\documentclass[prc,twocolumn,nofootinbib]{revtex4-2}

\usepackage{color}
\usepackage{graphicx}
\usepackage{dcolumn}
\newcolumntype{d}[1]{D{.}{.}{#1}}
\usepackage{bm}
\usepackage{braket}
\usepackage{amsmath}
\usepackage{here}
\usepackage{color}
\usepackage{multirow}

\usepackage{array}
\usepackage{enumerate}
\usepackage{amssymb}
\usepackage{bbm}
\usepackage{dsfont}
\usepackage[scr=rsfs]{mathalpha}
\usepackage{longtable}

\def\be{\begin{equation}} 
\def\ee{\end{equation}}

\usepackage{xcolor}
\graphicspath{{figs/}}
\usepackage[bookmarks=true,colorlinks,
            citecolor=blue,linkcolor=blue,anchorcolor=blue,filecolor=blue,urlcolor=blue,
           ]{hyperref}

\usepackage[normalem]{ulem}

\renewcommand\sout{\bgroup\markoverwith
{\textcolor[rgb]{1,0.75,0.8}{\rule[.5ex]{2pt}{0.8pt}}}\ULon}

\begin{document}

\title{
A microscopic calculation of fission cross sections 
with the non-equilibrium Green function method 
}

\author{K. Uzawa}
\thanks{Present address: Nuclear Data Center, Japan Atomic Energy Agency, Tokai, Ibaraki 319-1195, Japan.} 
\affiliation{Department of Physics, Kyoto University, Kyoto 606-8502, Japan}

\author{K. Hagino}
\affiliation{Department of Physics, Kyoto University, Kyoto 606-8502, Japan}

\begin{abstract}
We apply the non-equilibrium Green function (NEGF) method 
to microscopically 
evaluate fission cross sections for the neutron 
induced $^{235}$U$(n,f)$ reaction. 
While the model space 
was restricted only to seniority zero configurations 
in the previous applications of the NEGF method, 
we remove this restriction and include seniority non-zero 
configurations as well. In such model space, 
a proton-neutron 
interaction is active, for which we introduce a random interaction. We find that the seniority non-zero 
configurations significantly increase the fission 
cross sections, and thus the fission-to-capture branching 
ratios, even though they are still underestimated by about 
one order of magnitude 
as compared to the experimental data. 
In addition, we also find that 
the fission dynamics is governed by only a small 
number of eigenstates of the model Hamiltonian. 
\end{abstract}

\maketitle

\section{introduction}

Induced fission of heavy nuclei plays an important role in various phenomena, including r-process nucleosynthesis\cite{r-process1, r-process2, r-process3,goriely2015}  as well as syntheses of superheavy elements \cite{Hofmann2000,Hamilton2013, Giuliani2019,Hinde2021}. 
To describe these phenomena, 
it is thus crucial 
to understand deeply the fission process and make its reliable simulation. 
For that purpose, 
phenomenological approaches have often been applied.
The transition state theory \cite{Bohr1939, RevModPhys.62.251, truhlar1996current} is one of the good examples. In this theory, the existence of transition 
states is assumed at the saddle of a fission barrier 
and 
a fission decay rate is estimated by 
calculating the flux passing through the transition states. 
A diffusion model based on the Langevin dynamics \cite{Abe1996}
has also been widely used to describe shape evolution in 
the fission process \cite{Wada1993,Aritomo2013,Sierk2017, Ishizuka2017,Usang2019}. 
This method is particularly 
suitable to calculate mass and charge distributions 
of fission fragments.

Even though these phenomenological models have been 
successful, it is important to notice that they 
often rely on several 
theoretical assumptions, 
such as local thermal-equilibrium, whose 
applicability would have to be carefully examined 
when one would like to apply the models to unknown regions. 
For example, 
in r-process nucleosynthesis, 
the statistical approach may fail to describe 
the reaction process \cite{Mathews1983, Xu2014}, since 
neutron-rich nuclei at the end of the r-process path 
have a low neutron 
separation energy and thus the level density of the 
compound nuclei is also low. 
In such circumstances, a microscopic approach is more appropriate \cite{Schunck2016, Bender2020,Schunck2022}.
One of the promissing microscopic approaches is the time-dependent density functional theory (TDDFT) \cite{Nakatsukasa2016}, 
with which the understanding of the fission dynamics after the fission saddle point 
has been significantly advanced in recent years \cite{Bulgac2016}. In contrast to the post barrier dynamics, however, 
the pre-barrier dynamics, including a surmounting dynamics of a fission barrier and a connection to reaction 
theories, has not yet been completely clarified \cite{Bender2020}.

To develop a microscopic fission theory which is applicable both for the pre-barrier and the post-barrier dynamics, 
in this paper 
we particularly consider the non-equilibrium Green 
function (NEGF) approach \cite{Nonequilibrium_text, balzer2012nonequilibrium}. 
This method, combined with a microscopic Hamiltonian, 
has been widely used to calculate electron 
transport in nano-devices 
\cite{Datta1995,Datta2005,Nardelli1999,  Damle2001, Brandbyge2002, Xue2002,Camsari2023}. 
Regarding an induced fission as a transport phenomenon 
with shape evolution, 
the NEGF approach has been applied 
to induced fission reactions \cite{Bertsch2022, Weidenmuller2022, Bertsch2023, Uzawa2023, Uzawa2024}. 
A big challenge of this approach is that 
a fissioning nucleus in induced fission reactions 
is generally at a high excitation energy 
and the dimension of a Hamiltonian matrix is huge, which 
may reach the order of $10^5$ to $10^6$ \cite{Bertsch2023}. 
A direct computation of the Green function matrix, which requires inversion of the Hamiltonian matrix, therefore becomes numerically expensive \cite{Uzawa2024-2}. 
Because of this problem, 
the model space was restricted only to seniority zero 
configurations in the previous applications of the NEGF method to induced fission \cite{Bertsch2023, Uzawa2024}. 
For a realistic description of nuclear fission reactions, 
it is apparant that finite seniority 
configurations should also be included. 
The aim of this paper is to 
investigate the role of seniority non-zero configurations 
in induced fission by taking into account all possible 
configurations below a certain energy cut-off. 
Such study has been carried out for a schematic 
Hamiltonian \cite{Uzawa2023}, but has not yet been 
done for a realistic nucleus. 
We shall particularly analyze the $^{235}$U$(n,f)$ reaction at $E_n=10$ keV. 
We choose this energy because the $s$-wave dominantly contributes to the reaction around this energy
and the fission-to-capture branching ratio, $\alpha^{-1}$, 
varies only moderately \cite{Moore1984, Bertsch2017}. 
This will facilitate a comparison between the experimental data and the theoretical calculations.

The paper is organized as follows.
In Sec. \ref{Formulation},
we will detail the NEGF for induced fission and 
the setup of a model Hamiltonian based on the density 
functional theory. 
In Sec. \ref{Results}, we will present the results for the fission cross-sections in the $^{235}$U$(n,f)$ reaction and discuss the role of seniority non-zero configurations. 
In addition, 
by applying the spectrum decomposition of the Green function,
we will discuss the fission dynamics in terms of 
the eigenstates of the many-body Hamiltonian. 
Finally,  in Sec. \ref{summary}, we will
summarize the paper and discuss future perspectives.

\section{Formulation}\label{Formulation}

To carry out the NEGF calculation, one first has to prepare many-body basis functions 
covering model spaces which is large enough  to describe the fission process.
In this article, 
we generate basis functions with deformed Hartree-Fock calculations with 
the UNEDF1 Skyrme energy density functional \cite{Kortelainen2012}. 
The effective mass for this interaction is close to one, leading to  
a reasonable level density of excited states. 
We then compute the Hamiltonian and the overlap matrix elements among the basis states 
as is done in the generator coordinate method (GCM) \cite{ring}. 

\subsection{Fission path and model space}

\begin{figure}[htbp]
\centering
\includegraphics[width=8.6cm,clip]{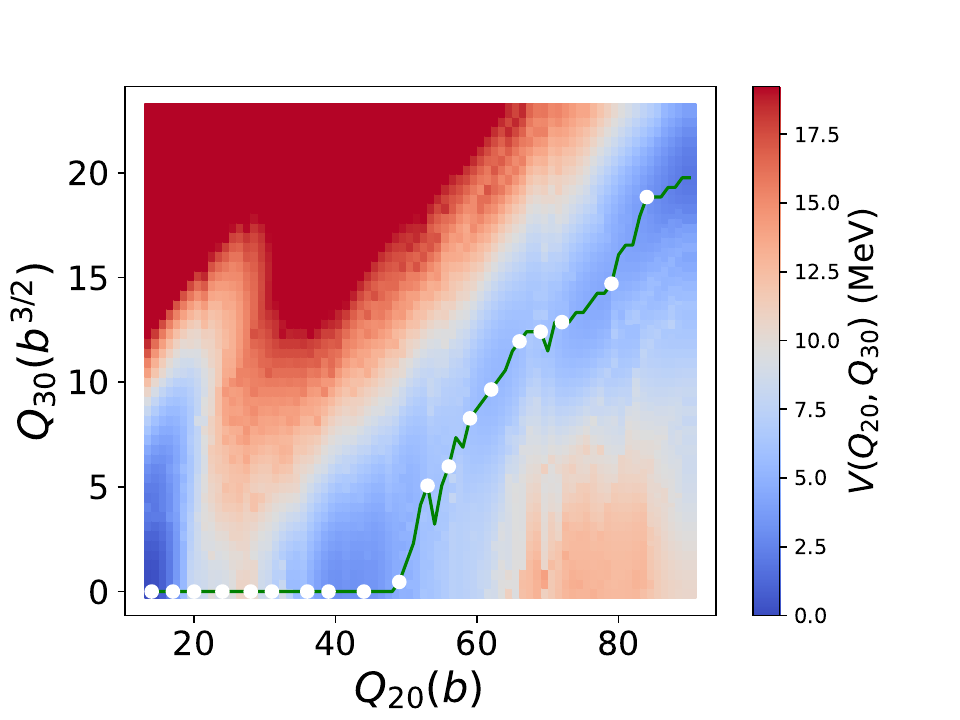}
\caption{
The potential energy surface of $^{236}$U as a function of the mass quadrupole and the octupole 
moments, $Q_{20}$ and $Q_{30}$, respectively, obtained with the Skyrme Hartree-Fock calculations with the UNEDF1 functional. 
The triaxial deformation and the pairing correlation are not taken into account. 
The green solid line shows the fission path determined as the valley of the potential energy surface, with the white points indicating the selected reference states used in the present calculations.
}
\label{PES}      
\end{figure}

\begin{figure}[htbp]
\centering
\includegraphics[width=8.6cm,clip]{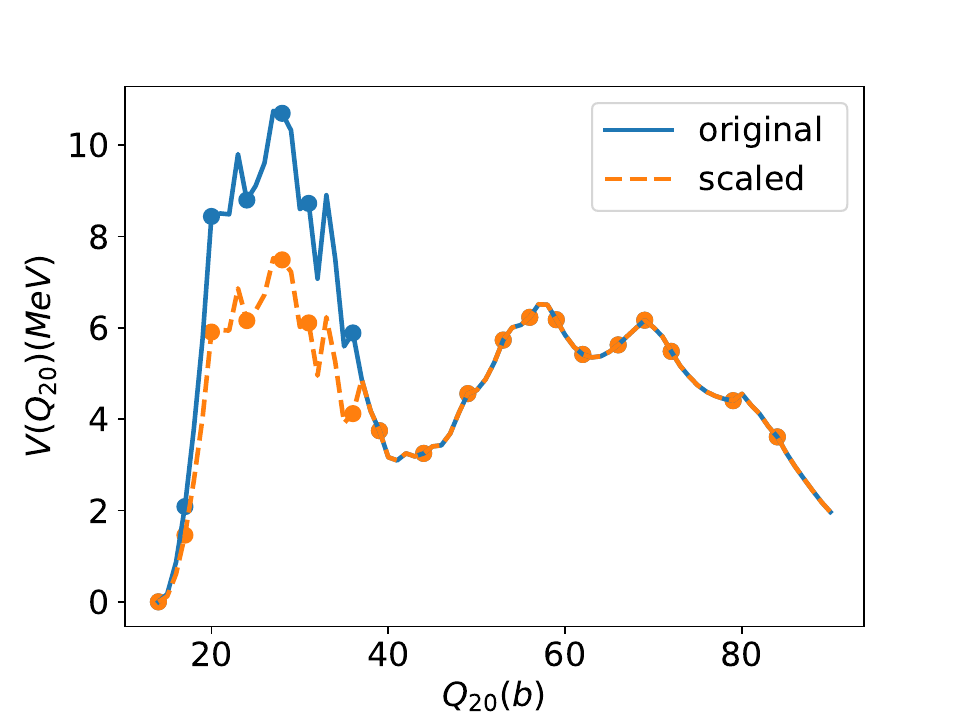}
\caption{
The fission barrier along the fission path shown in Fig. \ref{PES} as a function of the quadrupole moment $Q_{20}$.
The blue solid line shows the original barrier, while the orange dashed line is scaled by 
a factor of $f=0.7$. The points denote the reference states used for the GCM calculations.
}
\label{barrier}      
\end{figure}

We first calculate the potential energy surface (PES) and determine a fission 
path.
For this purpose, Fig. \ref{PES} shows the potential energy surface of  $^{236}$U as a function of the mass quadrupole 
moment $Q_{20}$ and the octupole moment $Q_{30}$ (to simplify the notation, 
in the following we shall use the notation $Q=Q_{20}$).
To draw the potential energy surface, 
we use the computer code {\tt SkyAx} \cite{Reinhard2021}, which assumes the axial symmetry
for nuclear shapes. 
The pairing interaction is not considered for a moment, even though we include it at the later stage of the calculations. 
We assume that the fission takes place along 
the valley of the PES, as shown by the green solid line in the figure.
The potential energy along the fission path, $V(Q)$, is shown 
by the blue solid line in Fig. \ref{barrier} as a function of the quadrupole moment $Q_{20}$.
Due to the lack of triaxial deformation and the pair correlation, 
the fission barrier height is overestimated as compared to the experimental data, 5.7 MeV \cite{Leal1999}. 
We therefore 
scale the first fission barrier by a factor of $f$. 
The value of $f$ is determined to reproduce the fission barrier height of 5.7 MeV 
after taking into account the residual pairing interaction, as we discuss in the next subsection. 
The orange dashed line in Fig. \ref{barrier} shows the scaled fission barrier with $f=0.7$. 

To construct the Green function to be used in the NEGF calculations, 
we discretize the fission path and calculate the Hamiltonian and the overlap integrals \cite{ring}. 
To this end, we set a criterion of the discretization to be  
$\langle Q_i|Q_{i+1}\rangle=0.52$ for the neigboring reference states, and discretize the fission path from 
$Q=14 ~{\rm b}$, where the ground state locates, to 
$Q=84 ~{\rm b}$ beyond the second fission barrier. 
Notice that the value for the criterion, 0.52,   
is somewhat larger than the 
value used in the previous calculations\cite{Bertsch2023, Uzawa2024}, that is, $1/e=0.36$, and hence 
the reference states are spaced more closely to each other.  
Even though the previous value was sufficient to get converged results for transmission coefficients, we 
use the larger value to obtain the converged ground state energy when the Hill-Wheeler equation is solved 
with the selected reference states. 
The selected reference states so obtained are shown by the dots in Fig. \ref{barrier}.

\begin{figure}[tbp]
\centering
\includegraphics[width=8.6cm,clip]{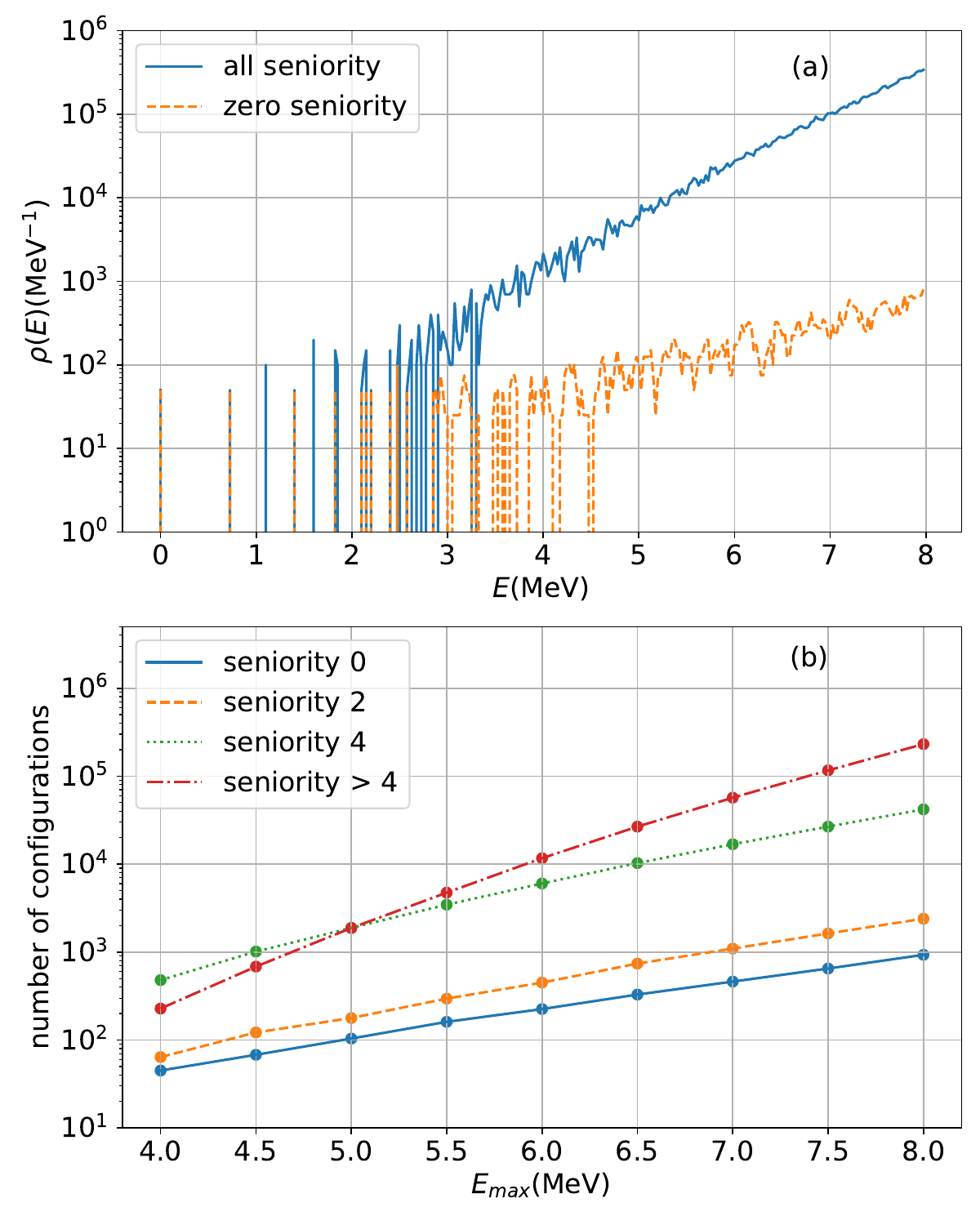}
\caption{
(a) 
The level densities $\rho(E)$ for $K=0$ configurations at $Q=14$ b. 
The blue-solid line shows the total level densities while the orange-dashed line is obtained 
by restricting to the zero seniority configurations only. 
(b)
The number of $K=0$ configurations at $Q=14 ~{\rm b}$ as a function of the energy cut-off, $E_{\rm max}$, for several seniority numbers. 
}
\label{seniority}      
\end{figure}

\begin{figure}[tbp]
\centering
\includegraphics[width=8.6cm,clip]{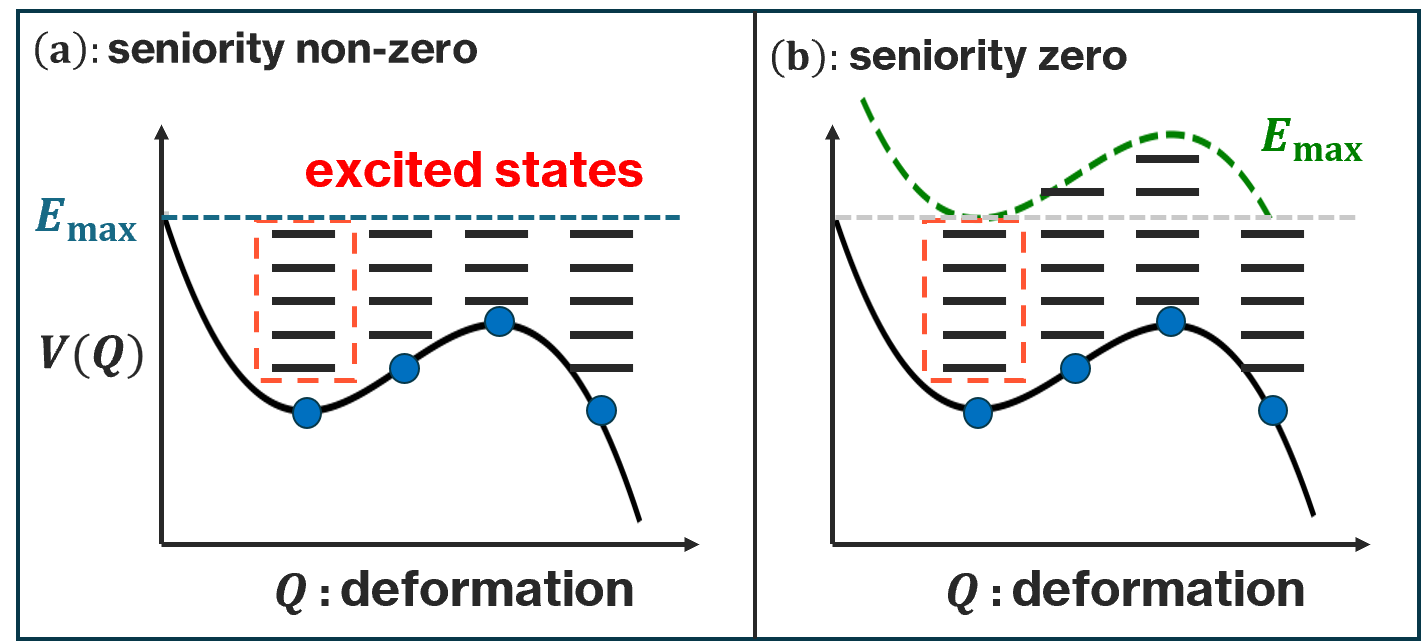}
\caption{
A schematic illustrations of the model space 
defined by Eqs. (\ref{cut1}) and (\ref{cut2}). 
The dashed lines indicate the energy cutoff for particle-hole excited states.
}
\label{model}      
\end{figure}

After the reference states are specified, we generate 
particle-hole excited configurations at each $Q$ using the single-particle levels at that deformation. 
Since we consider the axial symmetric deformations, we take into account only 
the configurations with the total $K$-quantum number of zero, where $K$ is the projection of angular momentum on 
the symmetry axis. 
The upper and the lower panels of Fig. \ref{seniority} show the level density $\rho$ and the number of configurations at $Q=14$ b, respectively. 
In both cases, the $K$-quantum number is restricted to $K=0$ only. 
The level density is plotted as a function of the excitation energy $E$ while the 
number of configurations is plotted as a function of the energy cut-off $E_{\rm max}$ 
for various values of the seniority quantum number. 
A typical excitation energy of the low-energy neutron-induced reaction of $^{235}$U 
is from 6 to 8 MeV, in which region the contribution of configurations with seniority 4 or larger 
becomes dominant. 
In the calculations shown below, 
we set up the energy cut-off for a configuration $|Q,E_\mu\rangle$ 
at $Q$ and the excitation energy of $E_\mu$ as (see Fig. \ref{model}),  
\begin{eqnarray}
E_\mu+V(Q) \le E_{\rm max} \  &:& {\rm finite \ seniority},\label{cut1}\\
E_\mu \le E_{\rm max} \ &:&{\rm zero \ seniority}.  \label{cut2} 
\end{eqnarray}
Notice that we take effectively a larger cut-off for the seniority zero configurations, 
in order to take into account a coherence of such configurations due to the pairing interactions. 
This choice is also convenient in determining the scaling factor for the fission barrier. 
We will specify the actual value of $E_{\rm max}$ in Sec. \ref{Results} after we will discuss 
the convergence of the results with respect to $E_{\rm max}$.  

\subsection{Hamiltonian and overlap integrals}\label{HandN}

For the interactions among the configurations specified in the previous subsection, 
we consider the monopole pairing interaction,
\begin{equation}
  H_{\rm pair}=-G\sum_{i\neq j}a^{\dagger}_{i}a^{\dagger}_{\bar{i}}a_{\bar{j}}a_{j},  
  \label{eq:H_pair}
\end{equation}
as well as the diabatic interaction \cite{Hagino2022},
\begin{eqnarray}
\frac{\langle Q,E_\mu|v_{db}|Q',E_{\mu'}\rangle}{\langle Q,E_\mu|Q',E_{\mu'}\rangle}
&=&\frac{E(Q,E_\mu)+E(Q',E_{\mu'})}{2} \nonumber \\
&&+h_2{\rm ln}\,(\langle Q,E_\mu|Q',E_{\mu'}\rangle). 
\label{eq:H_db}
\end{eqnarray}
Here, $\bar{i}$ is the time-reversed state of $i$. 
Those interactions have already been taken into accout in the previous works with 
seniority zero configurations only \cite{Bertsch2023,Uzawa2024}. 
In addition, 
we also introduce a random particle-hole interaction,
\begin{equation}
    H_{\rm ran}=vr\sum' a^\dagger_ia^\dagger_ja_la_k,
    \label{random}
\end{equation}
where the sum is restricted only to the configurations with the same $K$ quantum number 
and $r$ is a random number sampled from the standard normal distribution. 
Due to its stochastic nature, one has to repeat the calculations many times
and take an ensemble average of the results. 
This interaction 
originates from a contact type residual interaction\cite{Uzawa2023}, and its proton-neutron part acts only 
when the seniority non-zero 
configurations are taken into account. 
Since such interaction is known to play an important role in fission decay rates \cite{Bush1992}, 
and this is one of our motivations to go beyond the seniority zero approximation.

\begin{table}[tb] 
\caption{
\label{table:Gff} The values of the strength of the monopole pairing interaction, $G$, the scaling factor 
for the fission barrier, $f$, 
and the scaling factor $f_{\rm GOE}$ in Eq. (\ref{f_GOE}) 
with different values of the energy cut-off, $E_{\rm max}$.
}
\centering
\begin{tabular}{c|c|c|c}
  \hline
  \hline
  $E_{\rm max}$(MeV) & $G$ (MeV) & $f$   & $f_{\rm GOE}$ \\
  \hline
  6.0  & 0.1425  & 0.6810   &  35070\\
  6.5  &  0.1448 & 0.6905  &  23954 \\
  7.0  & 0.1465  &  0.6980 & 28288   \\
  7.5  &  0.1436 & 0.6973   & 26476   \\
  8.0  &  0.1412 & 0.6949   & 40964  \\
  \hline
  \hline
\end{tabular}
\end{table}

The value of $G$ in Eq. (\ref{eq:H_pair}) is determined to reproduce the excitation energy of 
the first excited $0^+$ state of $^{236}$U within a given model space with $E_{\rm max}$. 
The value of $G$ so obtained is summarized in Table \ref{table:Gff} for several values of $E_{\rm max}$.
The value of $f$ for the scaling factor of the fission barrier (see the previous subsection) 
is determined to reproduce the fission barrier height of 5.7 MeV 
after diagonalizing at each $Q$ the sub-Hamiltonian matrix with the residual pairing interaction.
The value of $f$ depends on the energy cut-off 
and its values are also summarized in Table \ref{table:Gff}. 
The average value of $h_2$ in Eq. (\ref{eq:H_db}) for the diabatic interaction 
was estimated to be $h_2=1.5$ MeV in Ref.\cite{Bertsch2023}, and we use the same value of $h_2$ in this paper.
The strength $v$ in Eq. (\ref{random}) for the random particle-hole interaction was microscopically 
estimated in Ref.\cite{Uzawa2023}.
Based on this estimation, 
we use $v=0.025$ MeV for the proton-neutron channel and $v=0.02$ MeV for the like-particle channels.

The three residual interactions provide off-diagonal matrix elements in the Hamiltonian overlaps, 
$H_{ii'}=\langle Q,E_\mu|\hat{H}|Q',E_{\mu'}\rangle$, where $|i\rangle$ is defined as $|i\rangle\equiv|Q,E_\mu\rangle$. 
Following Refs. \cite{Bertsch2023,Uzawa2024}, we take into account the couplings up to the nearest neighboring 
configurations with respect to the $Q$ coordinate. We also introduce the same approximation to the 
$N_{ii'}=\langle Q,E_\mu|Q',E_{\mu'}\rangle$ as well, whose off-diagonal 
components are set to be 0.52 
for the diabatically connected configurations and zero for the rest. 
Due to this approximation, the matrices $N$ and $H$ are block-tridiagonal, reducing the numerical costs to 
evaluate the Green function.
While in the previous publications 
the sub-Hamiltonian matrices at the left-most $Q$ and the right-most $Q$ were 
replaced by random matrices based on the Gaussian 
Orthogonal Ensemble (GOE) \cite{Bertsch2023,Uzawa2024}, in this paper we shall treat them as they are without introducing 
GOE matrices. 

\subsection{Decay width}\label{decay width}

In low-energy neutron induced reactions of actinide nuclei, 
neutron, capture, and fission channels are the dominant decay modes of a compound nucleus.
To take into account couplings to those decay channels, we add decay width matrices to the Hamiltonian matrix.
In our model, 
the configurations located at the left-end in the $Q$ coordinate, $Q=14 ~{\rm b}$, 
are considered to constitute the compound nucleus states, 
and are coupled to the neutron and the capture channels. 
On the other hand, the right-end configurations at $Q=84 ~{\rm b}$ are considered to be fission-doorway configurations 
which undergo fission. 
Therefore the decay width matrices read \cite{Uzawa2023},  
\begin{equation}
\label{gn}
    (\Gamma_n)_{ij}=\bar{\Gamma}_n \,N^{1/2}_{k_n,i}N^{1/2}_{k_n,j},
\end{equation}
\begin{equation}
\label{gc}
    (\Gamma_{\rm cap})_{ij}=\bar{\Gamma}_{\rm cap}\, \sum_{k\in Q_k=14 {\rm b}} N^{1/2}_{k,i}N^{1/2}_{k,j},
\end{equation}
and 
\begin{equation}
\label{gf}
    (\Gamma_{\rm fis})_{ij}=\bar{\Gamma}_{\rm fis} \sum_{k\in Q_k=84b} N^{1/2}_{k,i}N^{1/2}_{k,j},
\end{equation}
for the neutron, the capture, and the fission channels, respectively. 
Here $N_{ij}$ is the overlap integrals introduced in the previous subsection. 
Notice that 
the decay matrices were diagonal if the configurations were orthogonal to each other, that is $N_{ij}=\delta_{i,j}$. 
Eqs. (\ref{gn}-\ref{gf}) take into account the non-orthogonality of the configurations 
by introducing the square root of the overlap integrals. 
See Appendix B in Ref. \cite{Uzawa2023} for the derivation. 
$k_n$ denotes a single neutron channel which is coupled mainly to a selected single configuration at $Q=14 ~{\rm b}$. 
We select the neutron channel such that its diagonal energy in the Hamiltonian is the closest to the excitation 
energy $E$. 
In the energy average introduced in Sec. \ref{sec:cross_section} below,
we select
the configurations at $Q=14 ~{\rm b}$ 
whose excitation energy is within $[E-\Delta E/2, E+\Delta E/2]$,
and then calculate the cross-sections for each of them as the neutron entrance configuration. 
The energy average is defined as the average over those cross sections. 
The capture channels are also connected mainly to configurations at $Q=14 ~{\rm b}$. 
The number of the capture channels is empirically estimated to be 77 \cite{Leal1999}.
Thus we choose 
77 configurations at $Q=14 ~{\rm b}$ 
for $k$ in Eq. (\ref{gc}), with a criterion 
that the particle-hole excitation energies 
are closest to the total excitation energy $E$. 
Accordingly, we renormalize 
the capture width as $\bar{\Gamma}_{\rm cap}\rightarrow \bar{\Gamma}_{\rm cap}/77$.
On the other hand, the fission channels are connected to all the configurations at $Q=84~{\rm b}$.

\begin{figure}
\centering
\includegraphics[width=8.6cm]{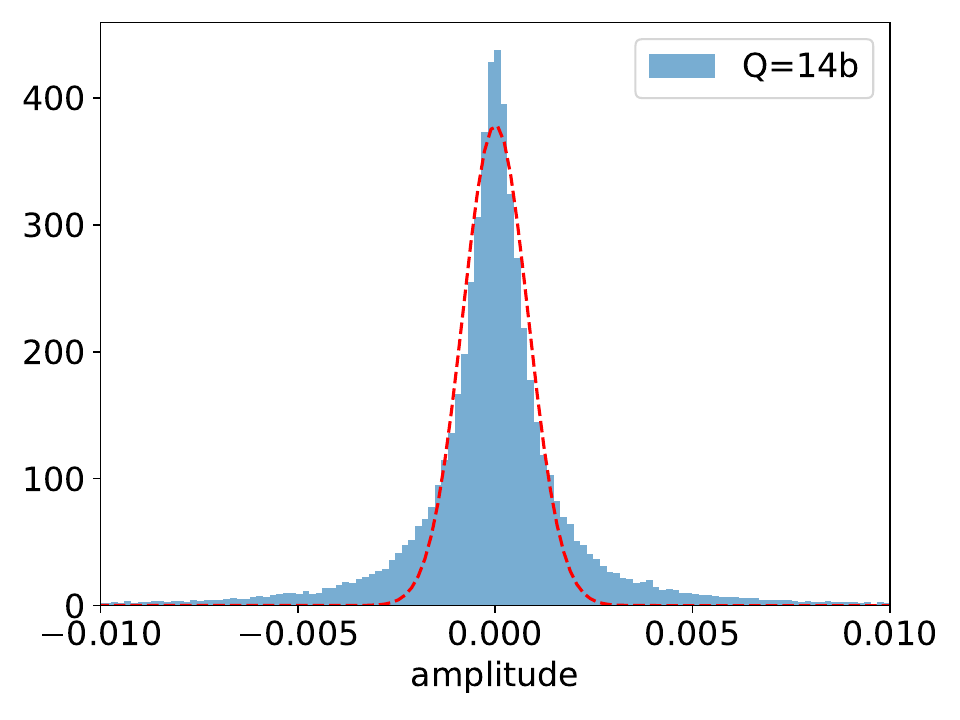}
\caption{The distribution of the expansion coefficients of an eigenvector at $Q=14$ b with $E_{\rm max}=7$ MeV. 
The eigenvector is chosen such that the corresponding eigenenergy is close to 6.546 MeV, that is, the 
excitation energy of the compound nucleus for an absorption of a thermal neutron. 
The red dashed line shows the result of a fitting to the Gaussian function.
}
\label{evec_Q14b}
\end{figure}

In Eqs. (\ref{gn}), (\ref{gc}), and (\ref{gf}), 
$\bar{\Gamma}_n, \bar{\Gamma}_{\rm cap}$, and $\bar{\Gamma}_{\rm fis}$ represent the sizes of the decay widths. 
The values of $\bar{\Gamma}_n$ and $\bar{\Gamma}_{\rm cap}$ can be determined from the nuclear data library RIPL \cite{ripl}, 
which summarizes reference values of transmission coefficients $T_n$ and $T_{\rm cap}$.  
For 
the $n+^{235}$U reaction at $E_n=10$ keV, one finds $T_n=0.0628$ and $T_{\rm cap}=0.25$ \cite{Bertsch2023}.
These values could be converted to average values of the decay widths using 
the compound nucleus phenomenology, 
\begin{equation}
\label{GOE_T}
T_i=2\pi\rho\bar{\Gamma}_i,     
\end{equation}
where $\rho$ is the level density of the compound nucleus. 
However, in our model, the compound nucleus assumption is not explicitly used, 
and this equation may not be directly applied. 
To test the validity of Eq. (\ref{GOE_T}), we diagonalize the first submatrix of $H_{ij}$ with $Q_i=Q_j=14~~{\rm b}$, 
and plot in Fig. \ref{evec_Q14b} the distribution of the component of the eigenvector 
whose eigenenergy is closest to the excitation energy of the compound nucleus for thermal neutrons, 6.546 MeV.
While the distribution is known to follow a Gaussian distribution for a GOE matrix \cite{Brody1981}, one can see 
that the distribution in our model significantly 
deviates from a Gauss distribution, probably because the Hamiltonian matrix in our model  
is much more sparse as compared to a GOE matrix due to the two-body nature of the residual interactions. 
To take into account this effect, 
we modify Eq.  (\ref{GOE_T}) by introducing a factor $f_{\rm GOE}$, that is,  
\begin{equation}
\label{f_GOE}
    T_i= 2\pi\rho\bar{\Gamma}_if_{\rm GOE}. 
\end{equation}
See Appendix A for a numerical method to determine 
the factor $f_{\rm GOE}$. 
The values of the decay widths are summarized in Table \ref{table:Gff} together with $f_{\rm GOE}$.

In contrast to $\bar{\Gamma}_{n}$ and $\bar{\Gamma}_{\rm cap}$,
the fission width $\bar{\Gamma}_{\rm fis}$ cannot be determined from 
experimental data. 
Also, its theoretical estimation has been limited so far \cite{Bertsch2019}. 
However, it has been shown numerically that 
calculated transmission coefficients in the NEGF approach 
were insensitive to the actual value of $\bar{\Gamma}_{\rm fis}$ \cite{Bertsch2023}. 
The insensitive property is a natural consequence of the fact that 
the fission probability is not affected by the post barrier dynamics 
if the back-scattered flux from the pre-fission configurations can be neglected. 
If this is the case, one can take an arbitrary value of $\bar{\Gamma}_{\rm fis}$ as long as it is large enough. 
We will check the validity of this assumption with the current setup 
in Sec. \ref{convergence with respects to}.

\begin{table}[tb] 
\caption{
\label{table:gamma} The sizes of the neutron width, $\bar{\Gamma}_n$, the capture width, $\bar{\Gamma}_{\rm cap}$, and the fission width, $\bar{\Gamma}_{\rm fis}$ with different values of the energy cut-off, $E_{\rm max}$. As explained in the main text, $\bar{\Gamma}_{\rm cap}$ is renormalized by dividing the number of capture channels, 77. }
\centering
\begin{tabular}{c|c|c|c}
  \hline
  \hline
  $E_{\rm max}$(MeV) & $\bar{\Gamma}_{n}$ (MeV) & $\bar{\Gamma}_{\rm cap}$ (MeV)  & $\bar{\Gamma}_{\rm fis}$ (MeV) \\
  \hline
  6.0  & 1.46$\times 10^{-2}$   & 7.47$\times 10^{-4}$  & 0.15  \\
  6.5  &  1.01$\times 10^{-2}$ & 5.23$\times 10^{-4}$  &  0.15 \\
  7.0  & 1.44$\times 10^{-2}$ &  7.46$\times 10^{-4}$ & 0.15  \\
  7.5  &  1.38$\times 10^{-2}$ & 7.14$\times 10^{-4}$  & 0.15  \\
  8.0  &  1.94$\times 10^{-2}$ & 1.01$\times 10^{-3}$  & 0.15 \\
  \hline
  \hline
\end{tabular}
\end{table}
\subsection{Green function and fission cross section}

Using the non-equilibrium Green function theory \cite{Caroli1971, Meir1992} 
or the S-matrix dispersion formula\cite{Alhassid2021},
the (retarded) Green function $G(E)$ defined by 
\begin{equation}
\label{green}
    G(E)=\left[EN-\left(H-\frac{i}{2}\Gamma\right)\right]^{-1}, 
\end{equation}
is related to the transmission coefficients $T_{ab}$ for a process 
from a channel $a$ to a channel $b~(\neq a)$ as, 
\begin{equation}
    T_{ab}={\rm Tr}[\Gamma_aG(E)\Gamma_bG^\dagger(E)], 
    \label{Tab}
\end{equation}
where $\Gamma_a$ and $\Gamma_b$ are the partial widths for the channels $a$ and $b$, respectively, 
and $\Gamma=\sum_a\Gamma_a$ in Eq. (\ref{green}) is the total width. 
In general a coupling to continuum channels leads to the self-energy terms, which have both the real and imaginary parts.
We simply neglect the real part and 
include only the imaginary part in Eqs. (\ref{green}) and (\ref{Tab}), provided that 
the uncertainty of the employed energy functional would be 
larger than the real part of the self-energy. 
For low-energy neutron reactions, only the $s$-wave component predominantly contributes and fission cross sections 
are given by, 
\begin{equation}
    \sigma_{n,\rm{fis}}=\frac{\pi}{k^2}T_{n,\rm{fis}},
    \label{eq:sigma_fis}
\end{equation}
where $k$ is the wave length of the incoming neutron.
Notice that 
the $K$-quantum number is restricted to $K=0$ 
in the calculations presented in the paper. 
Eq. (\ref{eq:sigma_fis}) would be justified if the transmission coefficient $T_{n,\rm{fis}}$ does not significantly 
depend on the $K$-quantum number.

\begin{figure}[tb]
\centering
\includegraphics[width=8.6cm]{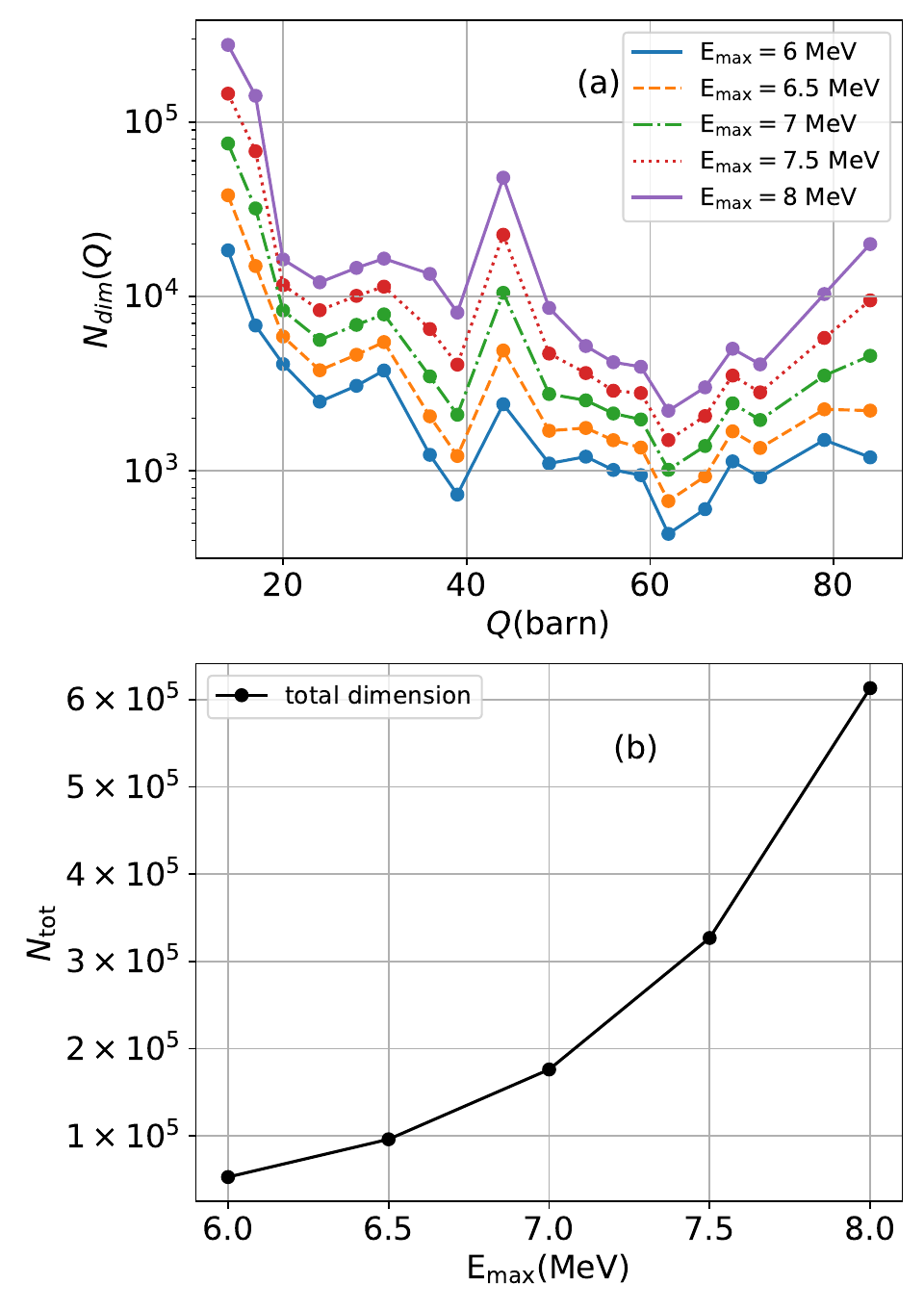}
\caption{
(a) The number of cofigurations at each $Q$ for different cut-off energies, $E_{\rm max}$.
(b) The total dimension of the whole Hamiltonian matrix, that is given as 
the sum of the number of configurations at each $Q$. 
}
\label{Ndim}
\end{figure}

In our model, as shown in Fig. \ref{Ndim}, 
the total dimension of the Hamiltonian matrix is $O(10^5)$, and 
a numerical cost for inverting the matrix in Eq. (\ref{green}) is huge. 
Notice that due to the random interaction, Eq. (\ref{random}), such calculation has to be repeated many times 
until the convergence is achieved. 
To overcome this problem, in Ref. \cite{Uzawa2024-2}, we have developed an efficient method based on 
the shift-invert Lanczos method. For the low-energy neutron reactions, one can also 
utilize the fact that the number of neutron channels is one \cite{Porter1956} and 
thus the transmission coefficient $T_{n,\rm{fis}}$ becomes
\begin{equation}
    T_{n,\rm{fis}}=\bar{\Gamma}_n\bar{\Gamma}_{\rm fis}\sum_{i\in \rm fis}|G_{i,n}|^2,
\end{equation}
if the overlap matrix $N$ is diagonal, i.e., $N=1$. 
In this case, it is sufficient to compute 
only one column of the Green function matrix for the channel $n$. 
That is, it only requires solving 
the simultaneous linear equations
\begin{equation}
\label{Gn_equation}
    \sum_j(EN-H)_{i,j}G_{j,n}=q_i, 
\end{equation}
where the vector $q$ is defined as $q_i=(0,\ldots,0,1,0,\ldots,0)^T$, 
having only one component for the channel $n$. 
This considerably reduces the numerical costs. 
In the actual calculations, 
the overlap matrix is not diagonal and the decay matrices are modified according to Eqs. (\ref{gn}-\ref{gf}).
Even in this case, 
one can still use a similar technique given that a configuration at $Q_i$ is connected in the 
overlap integral to a single configuration at 
the neighboring deformations, $Q=Q_{i\pm1}$, if the coupling is neglected for those which are 
not diabatically connected to each other.

We mension that 
the numerical cost is still large even with this numerical technique, 
and it is difficult to carry out the calculations with a standard supercomputer.
To achieve a realistic computation time, we apply the GPU version of the LSMR algorithm \cite{LSMR}, 
which in general shows good performance in solving simultaneous linear equations of sparse matrices. 
The usage of this algorithm and a high-performance GPU enable us to calculate the transmission 
coefficients $T_{n,{\rm fis}}$ with a large enough number of samples to take an ensemble average. 

\section{Results and discussion}\label{Results}

\subsection{Insensitivity properties}\label{convergence with respects to}

\begin{figure}
\centering
\includegraphics[width=8.6cm]{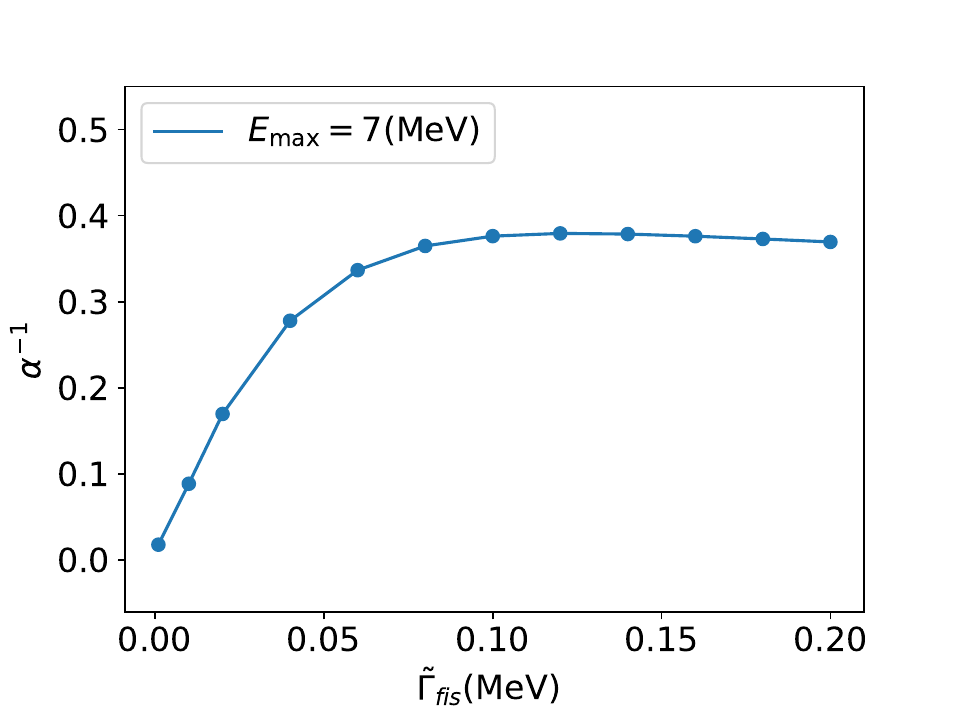}
\caption{
The fission-to-capture branching ratio, $\alpha^{-1}$, at $E=$ 6.546 MeV as a function of $\bar{\Gamma}_{\rm fis}$. 
The energy cut-off and the upper limit of $Q$ are set to be $E_{\rm max}=7$ MeV and $Q=84$ b, respectively.  
The result is obtained with a fixed random seed, without taking an ensemble and energry averages. 
}
\label{gf_insensitive}
\end{figure}

Let us now numerically evaluate fission cross sections. We first need to fix the value of $\bar{\Gamma}_{\rm fis}$. 
For this purpose,  
Fig. \ref{gf_insensitive} shows the fission-to-capture branching ratio, $\alpha^{-1}(E)=T_{n,{\rm fis}}(E)/T_{n,{\rm cap}}(E)$,  
with different values of $\bar{\Gamma}_{\rm fis}$ for the excitation energy of $E=6.546$ MeV.
As one can see, the branching ratio 
$\alpha^{-1}(E)$ converges with respect to $\bar{\Gamma}_{\rm fis}$ when $\bar{\Gamma}_{\rm fis}$ is large enough.  
While this figure shows the result with a single fixed random seed, 
we have confirmed that the conclusion remains the same with other random seeds. 
Therefore, in the following calculations, we use $\bar{\Gamma}_{\rm fis}=150$ keV,
at which the branching ratio converges as is seen in Fig. \ref{gf_insensitive}.

The insensitivity property shown in Fig. \ref{gf_insensitive} 
is similar to what has been found in the previous publications \cite{Bertsch2023,Uzawa2023}, 
implying that the flux passing through the fission barrier is not back-scattered from the pre-fission configurations \cite{Bertsch2023}. 
In this calculation, the upper limit of $Q$ is taken to be $Q=84$b, which is large enough to ensure a quasi-steady 
flow after the barrier, 
such that all the configurations at $Q=84$ b go into the fission decays.

\subsection{Fission cross section}
\label{sec:cross_section}

Our interest in this paper is to investigate 
whether the NEGF approach can quantitatively describe induced fission reactions. 
As one of the most important quantities in induced fission reactions is a fission cross section, 
we first discuss the convergence property of fission cross sections $\sigma_{n,\rm{fis}}(E)$ 
with respect to the cut-off energy, $E_{\rm max}$.
See Fig. \ref{Ndim} for the number of the configurations at each $Q$ and the dimension of the total Hamiltonian. 
The total size of the Hamiltonian matrix exponentially increases as $E_{\rm max}$ increases, reaching 
6$\times10^5$ at $E_{\rm max}=8$ MeV. 
To check the convergence of the fission cross sections, Eq. (\ref{eq:sigma_fis}), 
we fix the neutron energy to be $E_n=10$ keV, for which 
the excitation energy $E$ of $^{236}$U becomes $E=S_n+E_n=6.546$ MeV, 
with the empirical one-neutron separation energy $S_n=6.536$ MeV. 
Notice that the values of $G$, $f$ and $f_{\rm GOE}$ depend on $E_{\rm max}$, 
as shown in Table \ref{table:Gff}.

\begin{figure}
\centering
\includegraphics[width=8.6cm]{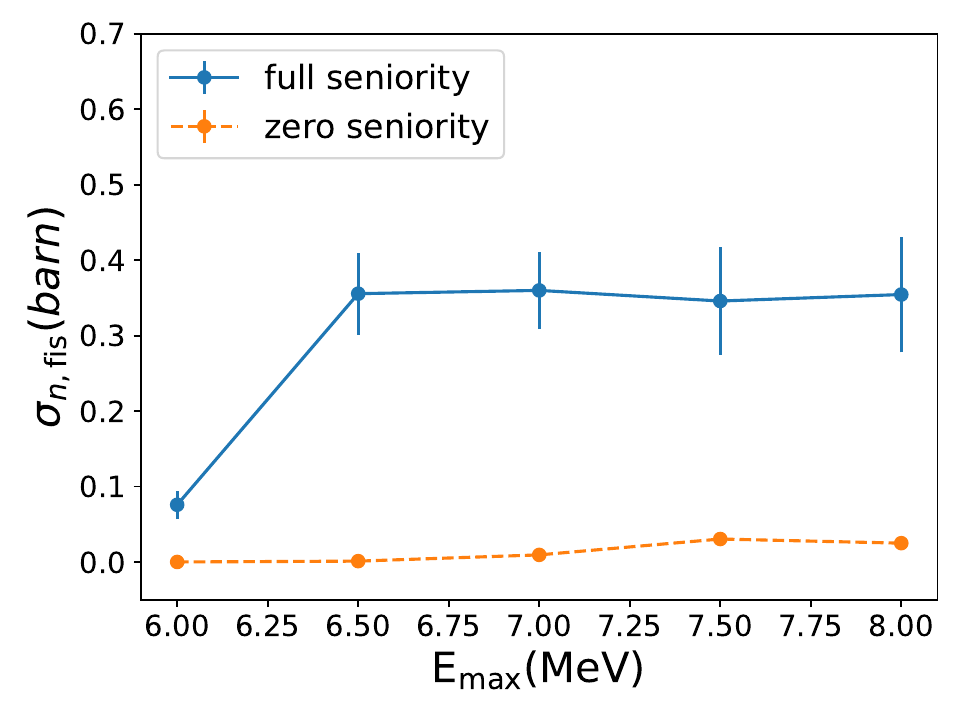}
\caption{
The averaged fission cross sections $\sigma_{n,{\rm fis}}$ at the excitation energy 
$E=6.546$ MeV
as a function of the cut-off energy, $E_{\rm max}$. 
The blue solid line shows the results with the full seniority configurations, for which the error bars are due to 
the statistical errors. The orange dashed line is obtained when only the seniority zero configurations are 
included in the calculations without changing the other parameters from the full seniority calculations. 
}
\label{sigma_full}
\end{figure}

The blue solid line in Fig. \ref{sigma_full} shows calculated fission cross sections for the $^{235}$U$(n,f)$ reaction. 
To this end, we take an energy average of the fission cross sections 
\begin{equation}
    \langle \sigma_{n, \rm fis}(E) \rangle=\frac{1}{\Delta E}\int^{E'+\Delta E/2}_{E'-\Delta E/2}dE' \sigma_{n, \rm fis}(E'),
\end{equation}
with $\Delta E=5$ keV. 
Due to the random interaction in the Hamiltonian, Eq. (\ref{random}), 
an ensemble average has to be taken also. 
For this purpose, we take 96 random seeds. 
In the figure, 
the error bars are due to the statistical errors.
Because of the huge numerical costs and the limitation of the computational resource, the remaining error 
is about 20\%.
Within the range of the error bars, one can see that 
the value of $\sigma_{n, \rm fis}$ is converged with respect to $E_{\rm max}$,
with $\sigma_{n, \rm fis}$=0.354 barn at $E_{\rm max}=8$ MeV.   
In the nuclear data library JENDL5.0 \cite{Iwamoto2023}, the fission cross section of $^{236}$U at $E$=6.546 MeV is 
$\sigma_{n,\rm fis}$=2.938 barn and thus our calculation underestimates the experimental data by 
about one order of magnitude. Notice that 
the maximum cross section at this energy would be $\pi/k^2=64.9$ barn, and thus a large portion of the 
total cross section is the compound elastic component. 
The orange dashed line in the figure shows the result 
when only the seniority zero configurations are included without changing the other parameters. 
This calculation substantially decreases the fission cross sections as compared to 
the full seniority calculations. 
This clearly indicates that the seniority non-zero configurations play an essential role 
in the fission cross sections.

\begin{figure}
\centering
\includegraphics[width=8.6cm]{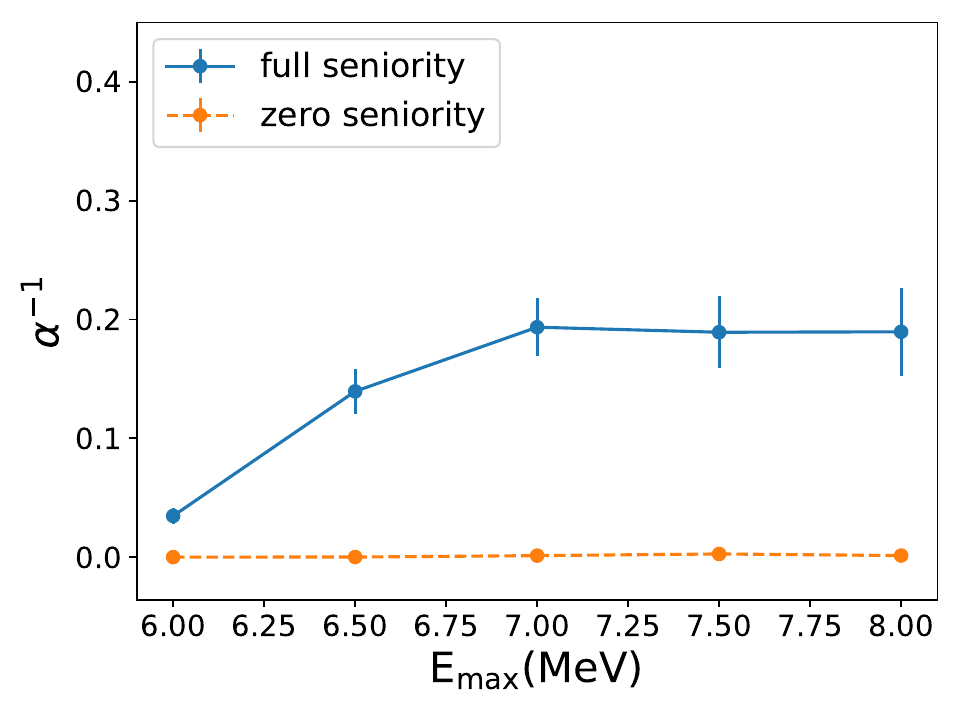}
\caption{
Same as Fig. \ref{sigma_full}, but for 
the average fission-to-capture branching ratio, $\alpha^{-1}$. 
}
\label{ratio_full}
\end{figure}

Fig. \ref{ratio_full} shows the corresponding fission-to-capture branching ratios,
\begin{equation}
    \langle \alpha^{-1}(E)\rangle=\frac{\int^{E'+\Delta E/2}_{E'-\Delta E/2}dE' \sigma_{n, \rm fis}(E')}{\int^{E'+\Delta E/2}_{E'-\Delta E/2}dE' \sigma_{n, \rm cap}(E')}.
\end{equation}
The behavior of $\alpha^{-1}(E)$ is qualitatively similar to $\sigma_{n, \rm fis}(E)$. 
That is, 
$\alpha^{-1}(E)$ is converged with respects to $E_{\rm max}$,
and is achieved $\alpha^{-1}(E)=0.190$, which if off from 
the nuclear library data, $\alpha^{-1}=2.831$ \cite{Moore1984}, by about one order of magnitude. 
Moreover, the seniority zero calculations largely underestimates the branching ratios, 
indicating once again the importance of the seniority non-zero configurations. 
Notice that 
the value of $\alpha^{-1}$ obtained with the seniority zero configurations is much smaller 
than the one obtained with the previous seniority zero calculations \cite{Bertsch2023,Uzawa2024}.  
In those previous works, the diagonal submatrix at $Q=14$ b was replaced by a GOE matrix, 
for which all the configurations are coupled to the configurations in the next $Q$ by a 
random interaction. The resultant 
off-diagonal submatrix connecting the GOE matrix 
and the Hamiltonian in the next $Q$ was dense. 
On the other hand, in the current setup, 
the off-diagonal submatrix is constructed based on the many-body Hamiltonian, and 
it is thus much more sparse. 
That is, 
the percentage of finite matrix elements is less than $1\%$. 
This considerably reduces the fission probability compared to the case with the dense off-diagonal submatrix, 
even though the size of the coupling strength is similar to each other. 

These converged results indicate that the model space is large enough if the cut-off energy is set 
to be $E_{\rm max}=7$ MeV or larger. 
Yet, the fission cross sections and thus the branching ratios are underestimated by about one order 
of magnitude. At this moment, the origin of this discrepancy is not clear. One possibility is 
that the triaxiality of nuclear shapes, which is not taken into account in the present study, may 
play some role. Another possibility is that the strength of the pairing interaction $G$ may have a large 
deformation dependence. Of course, the parameter set dependence of the energy functional would have to 
be checked also (see Appendix B for a related discussion). 
Inclusion of the momentum component \cite{Goeke1980,Hizawa2022,Hagino2024}
in the generator coordinate may also be important. 
All of these are beyond the scope of the 
present paper, and they would have to be examined in future applications of the NEGF method to induced fission 
reactions. In any case, the results obtained in this paper do not completely deviate from the empirical values 
and the NEGF approach provides a promising means to microscipically describe induced fission reactions. 

\subsection{Spectrum decomposition of transmission coefficients}

To gain a deeper insight into the fission dynamics, 
we decompose the transmission coefficients $T_{n,\rm fis}$ into the contribution from each of the eigenstates 
of the Hill-Wheeler equation \cite{ring},
\begin{eqnarray}
\left(H - \frac{i}{2} \Gamma\right) f_\lambda &=& \tilde{E}_\lambda N f_\lambda, \\
\label{eq:HW}
\left(H + \frac{i}{2} \Gamma\right) \tilde{f}_\lambda &=& \tilde{E}^*_\lambda N \tilde{f}_\lambda, 
\end{eqnarray}
with $\tilde{E}_\lambda\equiv E_\lambda - i\Gamma_\lambda/2$.
Using the eigenvectors $f_\lambda$ and $\tilde{f}_\lambda$
the Green function (\ref{green}) can be decomposed as,
\begin{equation}
\label{eq:G}
G(E)_{ij}
=\sum_\lambda \frac{\langle i|f_\lambda\rangle\langle \tilde{f}_\lambda|j\rangle}{E-\tilde{E}_\lambda} 
=\sum_\lambda \frac{f_\lambda(Q,E_\mu)\tilde{f}^*_\lambda(Q',E_{\mu'})}{E-\tilde{E}_\lambda}, 
\end{equation}
with $i=(Q,E_\mu)$ and $j=(Q',E_{\mu'})$. 
Note that the total Hamiltonian on the left hand side of Eq. (\ref{eq:HW}) 
is not hermitian due to the decay width term $i\Gamma/2$. 
In this case,  in general, 
the left-eigenvector $f_\lambda$ and the right-eigenvector $\tilde{f}_\lambda$ are not conjugate to each 
other.
However, in the problem of induced fission of actinide nuclei, 
the non-Hermite term $i\Gamma/2$ is much smaller 
than the hermitian term $H$,
and one can treat the non-hermitian part perturbatively. 
That is, in Eq.(\ref{eq:G}),
the same unperturbed eigenfunction $f^{(0)}_\lambda$ defined as 
\begin{equation}
H f^{(0)}_\lambda = E_\lambda N f^{(0)}_\lambda, 
\end{equation}
can be used both for $f_\lambda$ and $\tilde{f}_\lambda$, 
together with the imaginary part 
of the eigenenergy evaluated perturbatively as 
\begin{equation}
    \Gamma_\lambda=\sum_{i,j}(f^{(0)}_\lambda)^*_i \Gamma_{i,j} (f^{(0)}_\lambda)_j.
\end{equation}

Substituting the decomposed Green function (\ref{eq:G}) into Eq. (\ref{Tab}),
the transmission coefficients $T_{n,\rm fis}$ read
\begin{align}
\label{SD2}
&T_{n,\rm fis}
= \bar{\Gamma}_n\bar{\Gamma}_{\rm fis} \left[ \sum_\lambda  
\frac{|g_\lambda(Q_L,E_n)|^2 \left( \sum_{j \in {\rm fis}} |g_\lambda(Q_R,E_j)|^2 \right)}
{(E-E_\lambda)^2+(\Gamma_\lambda/2)^2} \right. \notag \\
&+ \left. \sum_{\lambda\neq\lambda'}  
\frac{G_{\lambda\lambda'}^{(n)}\,G_{\lambda\lambda'}^{({\rm fis})}}
{(E-E_\lambda+i\Gamma_\lambda/2)(E-E_{\lambda'}+i\Gamma_{\lambda'}/2)^*} \right],  
\end{align}
with
\begin{equation}
G_{\lambda\lambda'}^{(n)} \equiv g_\lambda(Q_L,E_n) g^*_{\lambda'}(Q_L,E_n), 
\end{equation}
and
\begin{equation}
G_{\lambda\lambda'}^{({\rm fis})}\equiv
\sum_{j \in {\rm fis}} g^*_\lambda(Q_R,E_j) g_{\lambda'}(Q_R,E_j).
\end{equation}
Here $g_\lambda$ is the collective wavefunction defined as \cite{ring},
\begin{equation}
    g_\lambda=N^{1/2}f^{(0)}_\lambda, 
\end{equation}
and 
$Q_L=14 ~{\rm b}$ and $Q_R=84 ~{\rm b}$ denote the left-most and the right-most configurations in the $Q$ coordinate, 
respectively (see Fig.\ref{barrier}).
The label $(Q_L,E_n)$ corresponds to the single configuration connected to the neutron channel $n$, while 
and the configurations labeled by $(Q_R,E_j)$ are connected to the fission channels.

\begin{figure}
\centering
\includegraphics[width=8.6cm]{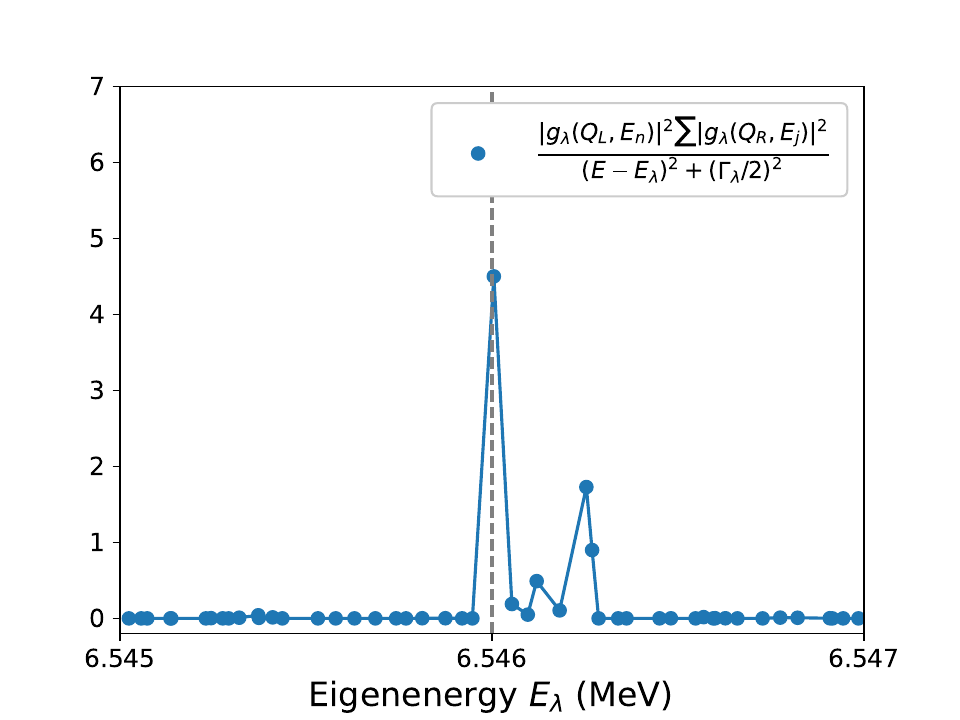}
\caption{
The first term on the right hand side of 
Eq. (\ref{SD2}) for  $E_{\rm max}=7$ MeV and $E=6.546$ MeV with a single random seed. 
It is plotted as a function of the 
eigenenergies $E_\lambda$ of the Hamiltonian. 
}
\label{SD_fig}
\end{figure}

The first term of the right hand side of Eq. (\ref{SD2}) 
is plotted in Fig. \ref{SD_fig} for $E$= 6.546 MeV and $E_{\rm max}=7$ MeV 
with a single random seed. It is plotted as a function of the 
real part of the eigenenergies, $E_\lambda$, 
obtained with the shift-invert Lanczos method to solve the large scale eigenvalue problem \cite{Shift-Invert, Uzawa2024-2}.
One can clearly see that 
only a few eigenstates contribute significantly to the transmission 
coefficient, $T_{n,\rm fis}$.  
Notice that the real part of the eigenenergy of those states is 
close to the excitation energy, $E=6.546$ MeV.
Because of the Breit–Wigner term $1/[(E-E_\lambda)^2+(\Gamma_\lambda/2)^2]$ in Eq. (\ref{SD2}), 
the contribution from those eigenstates whose eigenenergy is far from $E$ is suppressed.
Moreover, the eigenstates have to have large enough amplitudes both 
at $Q_L$ and at $Q_R$ in order to make a significant contribution to the 
transmission coefficient. 
A similar argument can be applied to the second term on the right hand side of Eq. (\ref{SD2}) with $\lambda\neq\lambda'$,  
because of the similar functional structure to the first term. 
In this way, 
the present model naturally accounts for the small number of the degrees of freedom in the fission channel \cite{Porter1956} 
without explicitly introducing transition states \cite{Uzawa2024}.

\section{Summary and future perspectives}\label{summary}

We have formulated a theoretical fission model based on the NEGF method with the density functional theory.
In contrast to the previous works, 
we have taken into account not only the seniority zero configurations but also the seniority non-zero 
configurations in the model space. 
Such extension has permitted one to take into account a random-type residual interaction, 
whose proton-neutron part had been shown to play an important role in induced fission reactions. 
Even though the inclusion of seniority non-zero configurations 
requires huge numerical costs to construct the Green function, 
we have succeeded in performing numerical calculations to evaluate the fission cross sections 
$\sigma_{n, \rm fis}$ as well as the fission-to-capture branching ratios $\alpha^{-1}$.
We have shown that those quantities were converged with respect to the energy cut-off $E_{\rm max}$, 
indicating that the model space employed in this paper was large enough and 
thus the NEGF approach was successfully applied to the induced fission process $^{235}$U$(n,f)$ reaction. 
We have found that our calculations reproduced 
the nuclear library data for 
the fission cross sections and the branching ratios within on order of magnitude. 
Considering that microscopic calculations of fission cross-sections for realistic systems have been regarded as 
extremely challenging, 
the results obtained in this paper, which do not completely deviate from the empirical values, 
appear encouraging for future developments of microscopic fission models.

Clearly, there are many possible extensions of the present approach. 
Fistly, in this paper, we have limited our study only to 
the $^{235}$U$(n,f)$ reaction at $E_n=10$ keV due 
to the huge numerical costs and a limitation of computational resources. 
An obvious extension of the present calculations 
is to other target nuclides and different energy ranges.  
In addition, in this paper, we have considered only 
the mass asymmetric fission path as shown in Fig. \ref{PES}. 
Even though 
the contribution of the mass symmetric fission mode is minor in the $^{235}$U$(n,f)$ reaction, 
it would be important to include both mass symmetric and asymmetric fission paths 
in several other nuclei. 
For that purpose, one would need to take into account a bifurcation of the fission path 
in the potential energy surface, or one would have to take both  
$Q_{20}$ and $Q_{30}$ as generator coordinates. In this way, one can 
also calculate a fragment mass distribution. 
Finally, 
in this paper, 
we have assumed that the transmission coefficient $T_{n,\rm{fis}}$ does not depends on 
the $K$-quantum number and have carried out the calculations only for $K=0$.
A more consistent treatment would be 
to take an average of $T_{n,\rm{fis}}$ obtained for different values of $K$. 
In this regard, one may also carry out the angular momentum projection. 
With these improvements, the 
the NEGF fission model would become a more powerful and reliable tool to 
microscopically describe nuclear fission reactions.

\section*{Acknowledgments}

We thank G.F. Bertsch for useful discussions, a careful reading of the manuscript, and 
his continuous encouragement. 
This work was supported by JSPS KAKENHI
Grant Nos. JP23K03414 and JP23KJ1212. 
The numerical calculations were performed
through the use of SQUID at the Cybermedia Center, Osaka University.

\appendix

\section{Determination of the factor $f_{\rm GOE}$}

\begin{figure}
\centering
\includegraphics[width=8.6cm]{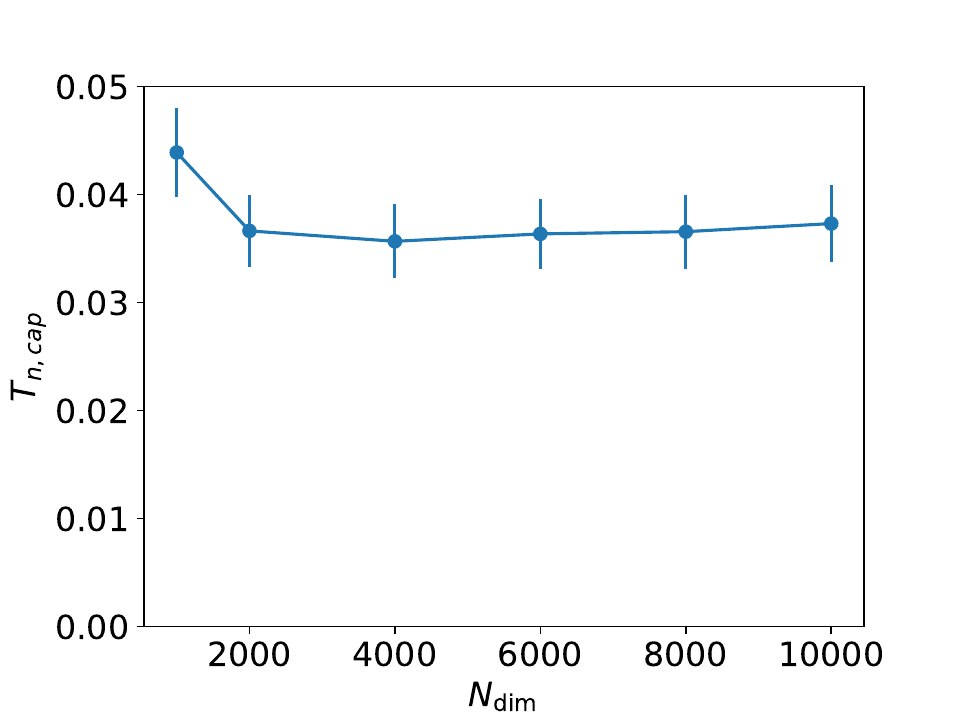}
\caption{
The transmission coefficient $T_{n,{\rm cap}}$ 
calculated with the GOE Hamiltonian as a function of the dimension of 
the matrix, $N_{\rm GOE}$. 
The parameter $v$ is determined to reproduced the level density, $\rho=1.96\times 10^4$. 
The decay widths in the Green function are scaled with $N_{\rm GOE}$ so that 
the result become independent of $N_{\rm dim}$ for large values of $N_{\rm dim}$.
}
\label{Tnc}
\end{figure}

We introduce the factor $f_{\rm GOE}$ in Eq. (\ref{f_GOE})  
to compensate the deviation of our model Hamiltonian from the compound nucleus models.
To describe compound nucleus reactions, one can assume that the GOE Hamiltonian $H_{\rm GOE}$ provides 
a good reference. 
The matrix elements of $H_{\rm GOE}$ are defined as 
\begin{equation}
    (H_{\rm GOE})_{ij}=vr\sqrt{1+\delta_{ij}}, 
\end{equation}
where $r$ is a random number sampled from the standard normal distribution, and $v$ is a 
scale of the matrix elements.
Using $H_{\rm GOE}$, we calculate the Green function at $E=6.546$ MeV, 
\begin{equation}
\label{G_GOE}
    G(E)=\left[E-H_{\rm GOE}+\frac{i}{2}(\Gamma_n+\Gamma_{\rm cap})\right]^{-1},
\end{equation}
with which the transmission coefficient $T_{n,{\rm cap}}$ can be obtained using the trace formula, Eq. (\ref{Tab}).
The decay width matrices in Eq. (\ref{G_GOE}) are defined by Eqs. (\ref{gn}) and (\ref{gc}), while 
the overlap matrix $N$ is set to be the identity matrix.
We estimate the decay widths $\bar{\Gamma}^{(0)}_n$ and $\bar{\Gamma}^{(0)}_{\rm cap}$ using the compound nucleus formula (\ref{GOE_T}) with the empirical values for $T_n$ and $T_{\rm cap}$, that is, 
$T_n=0.0628$ and  $T_{\rm cap}=0.025$ \cite{ripl, Bertsch2023}.
To this end, we use the level 
density $\rho$ determined from the energy spectrum at $Q=14$ b in the Hartree-Fock approximation, that is $\rho=1.96\times 10^4$ MeV$^{-1}$ for $E_{\rm max}=7$ MeV. 
The same value of  $\rho$ is used to determine the value of $v$ according to the relation in GOE, 
\begin{equation}
\label{rho_GOE}
    \rho=N_{\rm GOE}^{1/2}/\pi v, 
\end{equation}
for a given dimension $N_{\rm dim}$ of the GOE Hamiltonian. 
We scale $\bar{\Gamma}_i$ $(i=n \ {\rm or \ cap})$ as  $\bar{\Gamma}_i = N_{\rm dim}\times\bar{\Gamma}^{(0)}_i$ so that 
the value of $T_{n,{\rm cap}}$ becomes independent of $N_{\rm dim}$ as shown in Fig. \ref{Tnc}. 
We use the converged value $T_{n,{\rm cap}}=0.036$
as a reference to determine $f_{\rm GOE}$. The deviation of this value 
from the empirical value, $T_{n,{\rm cap}}=0.025$ indicates the deviation of 
our model Hamiltonian from the GOE model. 

\begin{figure}
\centering
\includegraphics[width=8.6cm]{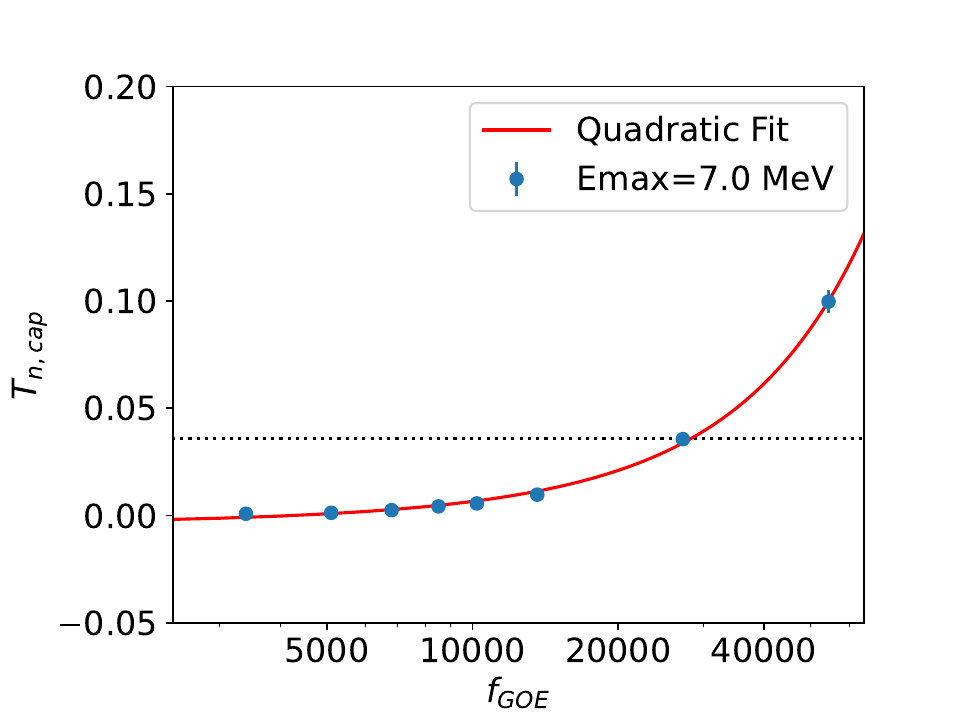}
\caption{
The transmission coefficient $T_{n,{\rm cap}}$ calculated 
with the Hamiltonian and the overlap matrices introduced in Sec. \ref{HandN} 
as a function of $f_{\rm GOE}$. 
The red solid line shows the result of a quadratic fitting. 
The dotted line indicates the converged value of $T_{n,{\rm cap}}=0.036$ 
shown in Fig.\ref{Tnc} obtained with the GOE model. 
}
\label{fGOE}
\end{figure}

With the decay widths $\bar{\Gamma}^{(0)}_n$ and $\bar{\Gamma}^{(0)}_{\rm cap}$ obtained in this way, 
we calculate $T_{n,{\rm cap}}$ using the model Hamiltonian 
introduced in Sec. \ref{HandN}. 
To this end, we set $\bar{\Gamma}_{\rm fis}$ to be zero.
Fig. \ref{fGOE} shows the result of $T_{n,{\rm cap}}$ with 
different $f_{\rm GOE}$, for each of which we scale $\bar{\Gamma}_n$ and $\bar{\Gamma}_{\rm cap}$ 
with $f_{\rm GOE}$, that is, 
$\bar{\Gamma}_i \to f_{\rm GOE}\times\bar{\Gamma}^{(0)}_i$. 
The energy cut-off is set to be $E_{\rm max}=7$ MeV. 
As one can see, $T_{n,{\rm cap}}$ can be well fitted by a 
quadric function of $f_{\rm GOE}$. 
The value of $f_{\rm GOE}$ is selected so that 
$T_{n,{\rm cap}}$ coincides with the value calculated with the GOE matrix, 
$T_{n,{\rm cap}}=0.036$. 

For each of $E_{\rm max}$, we repat the same procedure to determine $f_{\rm GOE}$, 
which are summarized in Table. \ref{table:Gff}.
Notice that the resultant values of $f_{\rm GOE}$ are $O(10^5)$. 
This is because this factor plays a role of the 
effective dimension of the compound nucleus configurations, similar to the case of 
the GOE Hamiltonian.  

\section{Parameter set dependence of the branching ratios}\label{AppendixE}

\begin{table}[tb] 
\caption{
\label{table:sky} 
The fission-to-capture branching ratio $\alpha^{-1}$ 
calculated with different Skyrme EDFs. 
The energy cut-off is set to be $E_{\rm max}=8$ MeV. 
The other parameters are the same as those in the main text. The effective mass $m^*/m$ 
and the total dimension of the Hamiltonian 
matrix $N_{\rm dim}$ are also tabulated. 
Notice that 
the calculation accuracy of double-precision numbers 
is about 16 decimal digits and the branching ratio for the SLy4 parameter 
set is consistent with zero. 
}
\centering
\begin{tabular}{c|ccc}
  \hline
  \hline
   & SLy4 & SkM*  & UNEDF1 \\
  \hline
  $m^*/m$  &  0.69 & 0.79  &  1.008 \\
  \hline
  $N_{\rm dim}$  &  73268 & 140886  &   613237 \\
  \hline
    $\alpha^{-1}$  & $1.94\times 10^{-18}$ & $1.91\times 10^{-4} $  & 0.190  \\
    \hline
    \hline
\end{tabular}
\end{table}

In the main text of this paper, we present the results with the UNEDF1 parameter set for the Skyrme-Hartree-Fock calculations. In this Appendix, 
we examine the dependence of the results on 
the parameter set of the Skyrme functional. 
Table \ref{table:sky} summarizes 
the fission-to-capture branching ratio $\alpha^{-1}$ obtained with three parameter sets, SLy4 \cite{CHABANAT1997710}, SkM*\cite{BARTEL198279} and UNEDF1. 
To this end, we set the cut-off energy to be $E_{\rm max}$ 
to be 8 MeV. 
One can see that 
$\alpha^{-1}$ is strongly correlated with 
the effective mass $m^*/m$, and decreases as the effective 
mass decreases. 
This is because a small effective mass results in 
a wide energy spacing of single-particle energy and 
thus reduces the level density. This is reflected in 
the number of total dimension, $N_{\rm dim}$. 
As a consequence, 
the transition from the initial state to the pre-fission configurations is largely suppressed. 
We therefore conclude that one needs to use a parameter 
set whose effective mass is close to unity 
for a microscopic description of induced fission.

\bibliography{references}

\begin{thebibliography}{62}%
\makeatletter
\providecommand \@ifxundefined [1]{%
 \@ifx{#1\undefined}
}%
\providecommand \@ifnum [1]{%
 \ifnum #1\expandafter \@firstoftwo
 \else \expandafter \@secondoftwo
 \fi
}%
\providecommand \@ifx [1]{%
 \ifx #1\expandafter \@firstoftwo
 \else \expandafter \@secondoftwo
 \fi
}%
\providecommand \natexlab [1]{#1}%
\providecommand \enquote  [1]{``#1''}%
\providecommand \bibnamefont  [1]{#1}%
\providecommand \bibfnamefont [1]{#1}%
\providecommand \citenamefont [1]{#1}%
\providecommand \href@noop [0]{\@secondoftwo}%
\providecommand \href [0]{\begingroup \@sanitize@url \@href}%
\providecommand \@href[1]{\@@startlink{#1}\@@href}%
\providecommand \@@href[1]{\endgroup#1\@@endlink}%
\providecommand \@sanitize@url [0]{\catcode `\\12\catcode `\$12\catcode `\&12\catcode `\#12\catcode `\^12\catcode `\_12\catcode `\%12\relax}%
\providecommand \@@startlink[1]{}%
\providecommand \@@endlink[0]{}%
\providecommand \url  [0]{\begingroup\@sanitize@url \@url }%
\providecommand \@url [1]{\endgroup\@href {#1}{\urlprefix }}%
\providecommand \urlprefix  [0]{URL }%
\providecommand \Eprint [0]{\href }%
\providecommand \doibase [0]{https://doi.org/}%
\providecommand \selectlanguage [0]{\@gobble}%
\providecommand \bibinfo  [0]{\@secondoftwo}%
\providecommand \bibfield  [0]{\@secondoftwo}%
\providecommand \translation [1]{[#1]}%
\providecommand \BibitemOpen [0]{}%
\providecommand \bibitemStop [0]{}%
\providecommand \bibitemNoStop [0]{.\EOS\space}%
\providecommand \EOS [0]{\spacefactor3000\relax}%
\providecommand \BibitemShut  [1]{\csname bibitem#1\endcsname}%
\let\auto@bib@innerbib\@empty
\bibitem [{\citenamefont {Kajino}\ \emph {et~al.}(2019)\citenamefont {Kajino}, \citenamefont {Aoki}, \citenamefont {Balantekin}, \citenamefont {Diehl}, \citenamefont {Famiano},\ and\ \citenamefont {Mathews}}]{r-process1}%
  \BibitemOpen
  \bibfield  {author} {\bibinfo {author} {\bibfnamefont {T.}~\bibnamefont {Kajino}}, \bibinfo {author} {\bibfnamefont {W.}~\bibnamefont {Aoki}}, \bibinfo {author} {\bibfnamefont {A.}~\bibnamefont {Balantekin}}, \bibinfo {author} {\bibfnamefont {R.}~\bibnamefont {Diehl}}, \bibinfo {author} {\bibfnamefont {M.}~\bibnamefont {Famiano}},\ and\ \bibinfo {author} {\bibfnamefont {G.}~\bibnamefont {Mathews}},\ }\bibfield  {title} {\bibinfo {title} {Current status of r-process nucleosynthesis},\ }\href {https://doi.org/https://doi.org/10.1016/j.ppnp.2019.02.008} {\bibfield  {journal} {\bibinfo  {journal} {Progress in Particle and Nuclear Physics}\ }\textbf {\bibinfo {volume} {107}},\ \bibinfo {pages} {109} (\bibinfo {year} {2019})}\BibitemShut {NoStop}%
\bibitem [{\citenamefont {Cowan}\ \emph {et~al.}(2021)\citenamefont {Cowan}, \citenamefont {Sneden}, \citenamefont {Lawler}, \citenamefont {Aprahamian}, \citenamefont {Wiescher}, \citenamefont {Langanke}, \citenamefont {Mart\'{\i}nez-Pinedo},\ and\ \citenamefont {Thielemann}}]{r-process2}%
  \BibitemOpen
  \bibfield  {author} {\bibinfo {author} {\bibfnamefont {J.~J.}\ \bibnamefont {Cowan}}, \bibinfo {author} {\bibfnamefont {C.}~\bibnamefont {Sneden}}, \bibinfo {author} {\bibfnamefont {J.~E.}\ \bibnamefont {Lawler}}, \bibinfo {author} {\bibfnamefont {A.}~\bibnamefont {Aprahamian}}, \bibinfo {author} {\bibfnamefont {M.}~\bibnamefont {Wiescher}}, \bibinfo {author} {\bibfnamefont {K.}~\bibnamefont {Langanke}}, \bibinfo {author} {\bibfnamefont {G.}~\bibnamefont {Mart\'{\i}nez-Pinedo}},\ and\ \bibinfo {author} {\bibfnamefont {F.-K.}\ \bibnamefont {Thielemann}},\ }\bibfield  {title} {\bibinfo {title} {Origin of the heaviest elements: The rapid neutron-capture process},\ }\href {https://doi.org/10.1103/RevModPhys.93.015002} {\bibfield  {journal} {\bibinfo  {journal} {Rev. Mod. Phys.}\ }\textbf {\bibinfo {volume} {93}},\ \bibinfo {pages} {015002} (\bibinfo {year} {2021})}\BibitemShut {NoStop}%
\bibitem [{\citenamefont {Mumpower}\ \emph {et~al.}(2016)\citenamefont {Mumpower}, \citenamefont {Surman}, \citenamefont {McLaughlin},\ and\ \citenamefont {Aprahamian}}]{r-process3}%
  \BibitemOpen
  \bibfield  {author} {\bibinfo {author} {\bibfnamefont {M.~R.}\ \bibnamefont {Mumpower}}, \bibinfo {author} {\bibfnamefont {R.}~\bibnamefont {Surman}}, \bibinfo {author} {\bibfnamefont {G.~C.}\ \bibnamefont {McLaughlin}},\ and\ \bibinfo {author} {\bibfnamefont {A.}~\bibnamefont {Aprahamian}},\ }\bibfield  {title} {\bibinfo {title} {The impact of individual nuclear properties on r-process nucleosynthesis},\ }\href {https://doi.org/https://doi.org/10.1016/j.ppnp.2015.09.001} {\bibfield  {journal} {\bibinfo  {journal} {Progress in Particle and Nuclear Physics}\ }\textbf {\bibinfo {volume} {86}},\ \bibinfo {pages} {86} (\bibinfo {year} {2016})}\BibitemShut {NoStop}%
\bibitem [{\citenamefont {Goriely}(2015)}]{goriely2015}%
  \BibitemOpen
  \bibfield  {author} {\bibinfo {author} {\bibfnamefont {S.}~\bibnamefont {Goriely}},\ }\bibfield  {title} {\bibinfo {title} {The fundamental role of fission during r-process nucleosynthesis in neutron star mergers},\ }\href@noop {} {\bibfield  {journal} {\bibinfo  {journal} {The European Physical Journal A}\ }\textbf {\bibinfo {volume} {51}},\ \bibinfo {pages} {1} (\bibinfo {year} {2015})}\BibitemShut {NoStop}%
\bibitem [{\citenamefont {Hofmann}\ and\ \citenamefont {M\"unzenberg}(2000)}]{Hofmann2000}%
  \BibitemOpen
  \bibfield  {author} {\bibinfo {author} {\bibfnamefont {S.}~\bibnamefont {Hofmann}}\ and\ \bibinfo {author} {\bibfnamefont {G.}~\bibnamefont {M\"unzenberg}},\ }\bibfield  {title} {\bibinfo {title} {The discovery of the heaviest elements},\ }\href {https://doi.org/10.1103/RevModPhys.72.733} {\bibfield  {journal} {\bibinfo  {journal} {Rev. Mod. Phys.}\ }\textbf {\bibinfo {volume} {72}},\ \bibinfo {pages} {733} (\bibinfo {year} {2000})}\BibitemShut {NoStop}%
\bibitem [{\citenamefont {Hamilton}\ \emph {et~al.}(2013)\citenamefont {Hamilton}, \citenamefont {Hofmann},\ and\ \citenamefont {Oganessian}}]{Hamilton2013}%
  \BibitemOpen
  \bibfield  {author} {\bibinfo {author} {\bibfnamefont {J.~H.}\ \bibnamefont {Hamilton}}, \bibinfo {author} {\bibfnamefont {S.}~\bibnamefont {Hofmann}},\ and\ \bibinfo {author} {\bibfnamefont {Y.~T.}\ \bibnamefont {Oganessian}},\ }\bibfield  {title} {\bibinfo {title} {Search for superheavy nuclei},\ }\href {https://doi.org/https://doi.org/10.1146/annurev-nucl-102912-144535} {\bibfield  {journal} {\bibinfo  {journal} {Annual Review of Nuclear and Particle Science}\ }\textbf {\bibinfo {volume} {63}},\ \bibinfo {pages} {383} (\bibinfo {year} {2013})}\BibitemShut {NoStop}%
\bibitem [{\citenamefont {Giuliani}\ \emph {et~al.}(2019)\citenamefont {Giuliani}, \citenamefont {Matheson}, \citenamefont {Nazarewicz}, \citenamefont {Olsen}, \citenamefont {Reinhard}, \citenamefont {Sadhukhan}, \citenamefont {Schuetrumpf}, \citenamefont {Schunck},\ and\ \citenamefont {Schwerdtfeger}}]{Giuliani2019}%
  \BibitemOpen
  \bibfield  {author} {\bibinfo {author} {\bibfnamefont {S.~A.}\ \bibnamefont {Giuliani}}, \bibinfo {author} {\bibfnamefont {Z.}~\bibnamefont {Matheson}}, \bibinfo {author} {\bibfnamefont {W.}~\bibnamefont {Nazarewicz}}, \bibinfo {author} {\bibfnamefont {E.}~\bibnamefont {Olsen}}, \bibinfo {author} {\bibfnamefont {P.-G.}\ \bibnamefont {Reinhard}}, \bibinfo {author} {\bibfnamefont {J.}~\bibnamefont {Sadhukhan}}, \bibinfo {author} {\bibfnamefont {B.}~\bibnamefont {Schuetrumpf}}, \bibinfo {author} {\bibfnamefont {N.}~\bibnamefont {Schunck}},\ and\ \bibinfo {author} {\bibfnamefont {P.}~\bibnamefont {Schwerdtfeger}},\ }\bibfield  {title} {\bibinfo {title} {{C}olloquium: Superheavy elements: Oganesson and beyond},\ }\href {https://doi.org/10.1103/RevModPhys.91.011001} {\bibfield  {journal} {\bibinfo  {journal} {Rev. Mod. Phys.}\ }\textbf {\bibinfo {volume} {91}},\ \bibinfo {pages} {011001} (\bibinfo {year} {2019})}\BibitemShut {NoStop}%
\bibitem [{\citenamefont {Hinde}\ \emph {et~al.}(2021)\citenamefont {Hinde}, \citenamefont {Dasgupta},\ and\ \citenamefont {Simpson}}]{Hinde2021}%
  \BibitemOpen
  \bibfield  {author} {\bibinfo {author} {\bibfnamefont {D.}~\bibnamefont {Hinde}}, \bibinfo {author} {\bibfnamefont {M.}~\bibnamefont {Dasgupta}},\ and\ \bibinfo {author} {\bibfnamefont {E.}~\bibnamefont {Simpson}},\ }\bibfield  {title} {\bibinfo {title} {Experimental studies of the competition between fusion and quasifission in the formation of heavy and superheavy nuclei},\ }\href {https://doi.org/https://doi.org/10.1016/j.ppnp.2021.103856} {\bibfield  {journal} {\bibinfo  {journal} {Progress in Particle and Nuclear Physics}\ }\textbf {\bibinfo {volume} {118}},\ \bibinfo {pages} {103856} (\bibinfo {year} {2021})}\BibitemShut {NoStop}%
\bibitem [{\citenamefont {Bohr}\ and\ \citenamefont {Wheeler}(1939)}]{Bohr1939}%
  \BibitemOpen
  \bibfield  {author} {\bibinfo {author} {\bibfnamefont {N.}~\bibnamefont {Bohr}}\ and\ \bibinfo {author} {\bibfnamefont {J.~A.}\ \bibnamefont {Wheeler}},\ }\bibfield  {title} {\bibinfo {title} {The mechanism of nuclear fission},\ }\href {https://doi.org/10.1103/PhysRev.56.426} {\bibfield  {journal} {\bibinfo  {journal} {Phys. Rev.}\ }\textbf {\bibinfo {volume} {56}},\ \bibinfo {pages} {426} (\bibinfo {year} {1939})}\BibitemShut {NoStop}%
\bibitem [{\citenamefont {H\"anggi}\ \emph {et~al.}(1990)\citenamefont {H\"anggi}, \citenamefont {Talkner},\ and\ \citenamefont {Borkovec}}]{RevModPhys.62.251}%
  \BibitemOpen
  \bibfield  {author} {\bibinfo {author} {\bibfnamefont {P.}~\bibnamefont {H\"anggi}}, \bibinfo {author} {\bibfnamefont {P.}~\bibnamefont {Talkner}},\ and\ \bibinfo {author} {\bibfnamefont {M.}~\bibnamefont {Borkovec}},\ }\bibfield  {title} {\bibinfo {title} {Reaction-rate theory: fifty years after kramers},\ }\href {https://doi.org/10.1103/RevModPhys.62.251} {\bibfield  {journal} {\bibinfo  {journal} {Rev. Mod. Phys.}\ }\textbf {\bibinfo {volume} {62}},\ \bibinfo {pages} {251} (\bibinfo {year} {1990})}\BibitemShut {NoStop}%
\bibitem [{\citenamefont {Truhlar}\ \emph {et~al.}(1996)\citenamefont {Truhlar}, \citenamefont {Garrett},\ and\ \citenamefont {Klippenstein}}]{truhlar1996current}%
  \BibitemOpen
  \bibfield  {author} {\bibinfo {author} {\bibfnamefont {D.~G.}\ \bibnamefont {Truhlar}}, \bibinfo {author} {\bibfnamefont {B.~C.}\ \bibnamefont {Garrett}},\ and\ \bibinfo {author} {\bibfnamefont {S.~J.}\ \bibnamefont {Klippenstein}},\ }\bibfield  {title} {\bibinfo {title} {Current status of transition-state theory},\ }\href@noop {} {\bibfield  {journal} {\bibinfo  {journal} {The Journal of physical chemistry}\ }\textbf {\bibinfo {volume} {100}},\ \bibinfo {pages} {12771} (\bibinfo {year} {1996})}\BibitemShut {NoStop}%
\bibitem [{\citenamefont {Abe}\ \emph {et~al.}(1996)\citenamefont {Abe}, \citenamefont {Ayik}, \citenamefont {Reinhard},\ and\ \citenamefont {Suraud}}]{Abe1996}%
  \BibitemOpen
  \bibfield  {author} {\bibinfo {author} {\bibfnamefont {Y.}~\bibnamefont {Abe}}, \bibinfo {author} {\bibfnamefont {S.}~\bibnamefont {Ayik}}, \bibinfo {author} {\bibfnamefont {P.-G.}\ \bibnamefont {Reinhard}},\ and\ \bibinfo {author} {\bibfnamefont {E.}~\bibnamefont {Suraud}},\ }\bibfield  {title} {\bibinfo {title} {On stochastic approaches of nuclear dynamics},\ }\href {https://doi.org/https://doi.org/10.1016/0370-1573(96)00003-8} {\bibfield  {journal} {\bibinfo  {journal} {Physics Reports}\ }\textbf {\bibinfo {volume} {275}},\ \bibinfo {pages} {49} (\bibinfo {year} {1996})}\BibitemShut {NoStop}%
\bibitem [{\citenamefont {Wada}\ \emph {et~al.}(1993)\citenamefont {Wada}, \citenamefont {Abe},\ and\ \citenamefont {Carjan}}]{Wada1993}%
  \BibitemOpen
  \bibfield  {author} {\bibinfo {author} {\bibfnamefont {T.}~\bibnamefont {Wada}}, \bibinfo {author} {\bibfnamefont {Y.}~\bibnamefont {Abe}},\ and\ \bibinfo {author} {\bibfnamefont {N.}~\bibnamefont {Carjan}},\ }\bibfield  {title} {\bibinfo {title} {One-body dissipation in agreement with prescission neutrons and fragment kinetic energies},\ }\href {https://doi.org/10.1103/PhysRevLett.70.3538} {\bibfield  {journal} {\bibinfo  {journal} {Phys. Rev. Lett.}\ }\textbf {\bibinfo {volume} {70}},\ \bibinfo {pages} {3538} (\bibinfo {year} {1993})}\BibitemShut {NoStop}%
\bibitem [{\citenamefont {Aritomo}\ and\ \citenamefont {Chiba}(2013)}]{Aritomo2013}%
  \BibitemOpen
  \bibfield  {author} {\bibinfo {author} {\bibfnamefont {Y.}~\bibnamefont {Aritomo}}\ and\ \bibinfo {author} {\bibfnamefont {S.}~\bibnamefont {Chiba}},\ }\bibfield  {title} {\bibinfo {title} {Fission process of nuclei at low excitation energies with a {L}angevin approach},\ }\href {https://doi.org/10.1103/PhysRevC.88.044614} {\bibfield  {journal} {\bibinfo  {journal} {Phys. Rev. C}\ }\textbf {\bibinfo {volume} {88}},\ \bibinfo {pages} {044614} (\bibinfo {year} {2013})}\BibitemShut {NoStop}%
\bibitem [{\citenamefont {Sierk}(2017)}]{Sierk2017}%
  \BibitemOpen
  \bibfield  {author} {\bibinfo {author} {\bibfnamefont {A.~J.}\ \bibnamefont {Sierk}},\ }\bibfield  {title} {\bibinfo {title} {{L}angevin model of low-energy fission},\ }\href {https://doi.org/10.1103/PhysRevC.96.034603} {\bibfield  {journal} {\bibinfo  {journal} {Phys. Rev. C}\ }\textbf {\bibinfo {volume} {96}},\ \bibinfo {pages} {034603} (\bibinfo {year} {2017})}\BibitemShut {NoStop}%
\bibitem [{\citenamefont {Ishizuka}\ \emph {et~al.}(2017)\citenamefont {Ishizuka}, \citenamefont {Usang}, \citenamefont {Ivanyuk}, \citenamefont {Maruhn}, \citenamefont {Nishio},\ and\ \citenamefont {Chiba}}]{Ishizuka2017}%
  \BibitemOpen
  \bibfield  {author} {\bibinfo {author} {\bibfnamefont {C.}~\bibnamefont {Ishizuka}}, \bibinfo {author} {\bibfnamefont {M.~D.}\ \bibnamefont {Usang}}, \bibinfo {author} {\bibfnamefont {F.~A.}\ \bibnamefont {Ivanyuk}}, \bibinfo {author} {\bibfnamefont {J.~A.}\ \bibnamefont {Maruhn}}, \bibinfo {author} {\bibfnamefont {K.}~\bibnamefont {Nishio}},\ and\ \bibinfo {author} {\bibfnamefont {S.}~\bibnamefont {Chiba}},\ }\bibfield  {title} {\bibinfo {title} {Four-dimensional {L}angevin approach to low-energy nuclear fission of $^{236}\mathbf{U}$},\ }\href {https://doi.org/10.1103/PhysRevC.96.064616} {\bibfield  {journal} {\bibinfo  {journal} {Phys. Rev. C}\ }\textbf {\bibinfo {volume} {96}},\ \bibinfo {pages} {064616} (\bibinfo {year} {2017})}\BibitemShut {NoStop}%
\bibitem [{\citenamefont {Usang}\ \emph {et~al.}(2019)\citenamefont {Usang}, \citenamefont {Ivanyuk}, \citenamefont {Ishizuka},\ and\ \citenamefont {Chiba}}]{Usang2019}%
  \BibitemOpen
  \bibfield  {author} {\bibinfo {author} {\bibfnamefont {M.~D.}\ \bibnamefont {Usang}}, \bibinfo {author} {\bibfnamefont {F.~A.}\ \bibnamefont {Ivanyuk}}, \bibinfo {author} {\bibfnamefont {C.}~\bibnamefont {Ishizuka}},\ and\ \bibinfo {author} {\bibfnamefont {S.}~\bibnamefont {Chiba}},\ }\bibfield  {title} {\bibinfo {title} {Correlated transitions in {TKE} and mass distributions of fission fragments described by 4-{D} {L}angevin equation},\ }\href@noop {} {\bibfield  {journal} {\bibinfo  {journal} {Scientific reports}\ }\textbf {\bibinfo {volume} {9}},\ \bibinfo {pages} {1525} (\bibinfo {year} {2019})}\BibitemShut {NoStop}%
\bibitem [{\citenamefont {{Mathews}}\ \emph {et~al.}(1983)\citenamefont {{Mathews}}, \citenamefont {{Mengoni}}, \citenamefont {{Thielemann}},\ and\ \citenamefont {{Fowler}}}]{Mathews1983}%
  \BibitemOpen
  \bibfield  {author} {\bibinfo {author} {\bibfnamefont {G.~J.}\ \bibnamefont {{Mathews}}}, \bibinfo {author} {\bibfnamefont {A.}~\bibnamefont {{Mengoni}}}, \bibinfo {author} {\bibfnamefont {F.~K.}\ \bibnamefont {{Thielemann}}},\ and\ \bibinfo {author} {\bibfnamefont {W.~A.}\ \bibnamefont {{Fowler}}},\ }\bibfield  {title} {\bibinfo {title} {{Neutron capture rates in the r-process - The role of direct radiative capture}},\ }\href {https://doi.org/10.1086/161164} {\bibfield  {journal} {\bibinfo  {journal} {\apj}\ }\textbf {\bibinfo {volume} {270}},\ \bibinfo {pages} {740} (\bibinfo {year} {1983})}\BibitemShut {NoStop}%
\bibitem [{\citenamefont {Xu}\ \emph {et~al.}(2014)\citenamefont {Xu}, \citenamefont {Goriely}, \citenamefont {Koning},\ and\ \citenamefont {Hilaire}}]{Xu2014}%
  \BibitemOpen
  \bibfield  {author} {\bibinfo {author} {\bibfnamefont {Y.}~\bibnamefont {Xu}}, \bibinfo {author} {\bibfnamefont {S.}~\bibnamefont {Goriely}}, \bibinfo {author} {\bibfnamefont {A.~J.}\ \bibnamefont {Koning}},\ and\ \bibinfo {author} {\bibfnamefont {S.}~\bibnamefont {Hilaire}},\ }\bibfield  {title} {\bibinfo {title} {Systematic study of neutron capture including the compound, pre-equilibrium, and direct mechanisms},\ }\href {https://doi.org/10.1103/PhysRevC.90.024604} {\bibfield  {journal} {\bibinfo  {journal} {Phys. Rev. C}\ }\textbf {\bibinfo {volume} {90}},\ \bibinfo {pages} {024604} (\bibinfo {year} {2014})}\BibitemShut {NoStop}%
\bibitem [{\citenamefont {Schunck}\ and\ \citenamefont {Robledo}(2016)}]{Schunck2016}%
  \BibitemOpen
  \bibfield  {author} {\bibinfo {author} {\bibfnamefont {N.}~\bibnamefont {Schunck}}\ and\ \bibinfo {author} {\bibfnamefont {L.~M.}\ \bibnamefont {Robledo}},\ }\bibfield  {title} {\bibinfo {title} {Microscopic theory of nuclear fission: a review},\ }\href {https://doi.org/10.1088/0034-4885/79/11/116301} {\bibfield  {journal} {\bibinfo  {journal} {Reports on Progress in Physics}\ }\textbf {\bibinfo {volume} {79}},\ \bibinfo {pages} {116301} (\bibinfo {year} {2016})}\BibitemShut {NoStop}%
\bibitem [{\citenamefont {Bender}\ \emph {et~al.}(2020)\citenamefont {Bender} \emph {et~al.}}]{Bender2020}%
  \BibitemOpen
  \bibfield  {author} {\bibinfo {author} {\bibfnamefont {M.}~\bibnamefont {Bender}} \emph {et~al.},\ }\bibfield  {title} {\bibinfo {title} {Future of nuclear fission theory},\ }\href {https://doi.org/10.1088/1361-6471/abab4f} {\bibfield  {journal} {\bibinfo  {journal} {Journal of Physics G: Nuclear and Particle Physics}\ }\textbf {\bibinfo {volume} {47}},\ \bibinfo {pages} {113002} (\bibinfo {year} {2020})}\BibitemShut {NoStop}%
\bibitem [{\citenamefont {Schunck}\ and\ \citenamefont {Regnier}(2022)}]{Schunck2022}%
  \BibitemOpen
  \bibfield  {author} {\bibinfo {author} {\bibfnamefont {N.}~\bibnamefont {Schunck}}\ and\ \bibinfo {author} {\bibfnamefont {D.}~\bibnamefont {Regnier}},\ }\bibfield  {title} {\bibinfo {title} {Theory of nuclear fission},\ }\href {https://doi.org/https://doi.org/10.1016/j.ppnp.2022.103963} {\bibfield  {journal} {\bibinfo  {journal} {Progress in Particle and Nuclear Physics}\ }\textbf {\bibinfo {volume} {125}},\ \bibinfo {pages} {103963} (\bibinfo {year} {2022})}\BibitemShut {NoStop}%
\bibitem [{\citenamefont {Nakatsukasa}\ \emph {et~al.}(2016)\citenamefont {Nakatsukasa}, \citenamefont {Matsuyanagi}, \citenamefont {Matsuo},\ and\ \citenamefont {Yabana}}]{Nakatsukasa2016}%
  \BibitemOpen
  \bibfield  {author} {\bibinfo {author} {\bibfnamefont {T.}~\bibnamefont {Nakatsukasa}}, \bibinfo {author} {\bibfnamefont {K.}~\bibnamefont {Matsuyanagi}}, \bibinfo {author} {\bibfnamefont {M.}~\bibnamefont {Matsuo}},\ and\ \bibinfo {author} {\bibfnamefont {K.}~\bibnamefont {Yabana}},\ }\bibfield  {title} {\bibinfo {title} {Time-dependent density-functional description of nuclear dynamics},\ }\href {https://doi.org/10.1103/RevModPhys.88.045004} {\bibfield  {journal} {\bibinfo  {journal} {Rev. Mod. Phys.}\ }\textbf {\bibinfo {volume} {88}},\ \bibinfo {pages} {045004} (\bibinfo {year} {2016})}\BibitemShut {NoStop}%
\bibitem [{\citenamefont {Bulgac}\ \emph {et~al.}(2016)\citenamefont {Bulgac}, \citenamefont {Magierski}, \citenamefont {Roche},\ and\ \citenamefont {Stetcu}}]{Bulgac2016}%
  \BibitemOpen
  \bibfield  {author} {\bibinfo {author} {\bibfnamefont {A.}~\bibnamefont {Bulgac}}, \bibinfo {author} {\bibfnamefont {P.}~\bibnamefont {Magierski}}, \bibinfo {author} {\bibfnamefont {K.~J.}\ \bibnamefont {Roche}},\ and\ \bibinfo {author} {\bibfnamefont {I.}~\bibnamefont {Stetcu}},\ }\bibfield  {title} {\bibinfo {title} {Induced fission of $^{240}\mathrm{Pu}$ within a real-time microscopic framework},\ }\href {https://doi.org/10.1103/PhysRevLett.116.122504} {\bibfield  {journal} {\bibinfo  {journal} {Phys. Rev. Lett.}\ }\textbf {\bibinfo {volume} {116}},\ \bibinfo {pages} {122504} (\bibinfo {year} {2016})}\BibitemShut {NoStop}%
\bibitem [{\citenamefont {Stefanucci}\ and\ \citenamefont {van Leeuwen}(2013)}]{Nonequilibrium_text}%
  \BibitemOpen
  \bibfield  {author} {\bibinfo {author} {\bibfnamefont {G.}~\bibnamefont {Stefanucci}}\ and\ \bibinfo {author} {\bibfnamefont {R.}~\bibnamefont {van Leeuwen}},\ }\href@noop {} {\emph {\bibinfo {title} {Nonequilibrium Many-Body Theory of Quantum Systems: A Modern Introduction}}}\ (\bibinfo  {publisher} {Cambridge University Press},\ \bibinfo {year} {2013})\BibitemShut {NoStop}%
\bibitem [{\citenamefont {Balzer}\ and\ \citenamefont {Bonitz}(2012)}]{balzer2012nonequilibrium}%
  \BibitemOpen
  \bibfield  {author} {\bibinfo {author} {\bibfnamefont {K.}~\bibnamefont {Balzer}}\ and\ \bibinfo {author} {\bibfnamefont {M.}~\bibnamefont {Bonitz}},\ }\href@noop {} {\emph {\bibinfo {title} {Nonequilibrium {G}reen's Functions Approach to Inhomogeneous Systems}}}\ (\bibinfo  {publisher} {Springer},\ \bibinfo {year} {2012})\BibitemShut {NoStop}%
\bibitem [{\citenamefont {Datta}(1995)}]{Datta1995}%
  \BibitemOpen
  \bibfield  {author} {\bibinfo {author} {\bibfnamefont {S.}~\bibnamefont {Datta}},\ }\href@noop {} {\emph {\bibinfo {title} {Electronic Transport in Mesoscopic Systems}}}\ (\bibinfo  {publisher} {Cambridge University Press, Cambridge},\ \bibinfo {year} {1995})\BibitemShut {NoStop}%
\bibitem [{\citenamefont {Datta}(2005)}]{Datta2005}%
  \BibitemOpen
  \bibfield  {author} {\bibinfo {author} {\bibfnamefont {S.}~\bibnamefont {Datta}},\ }\href@noop {} {\emph {\bibinfo {title} {Quantum Transport: Atom to Transistor}}}\ (\bibinfo  {publisher} {Cambridge University Press},\ \bibinfo {year} {2005})\BibitemShut {NoStop}%
\bibitem [{\citenamefont {Nardelli}(1999)}]{Nardelli1999}%
  \BibitemOpen
  \bibfield  {author} {\bibinfo {author} {\bibfnamefont {M.~B.}\ \bibnamefont {Nardelli}},\ }\bibfield  {title} {\bibinfo {title} {Electronic transport in extended systems: Application to carbon nanotubes},\ }\href {https://doi.org/10.1103/PhysRevB.60.7828} {\bibfield  {journal} {\bibinfo  {journal} {Phys. Rev. B}\ }\textbf {\bibinfo {volume} {60}},\ \bibinfo {pages} {7828} (\bibinfo {year} {1999})}\BibitemShut {NoStop}%
\bibitem [{\citenamefont {Damle}\ \emph {et~al.}(2001)\citenamefont {Damle}, \citenamefont {Ghosh},\ and\ \citenamefont {Datta}}]{Damle2001}%
  \BibitemOpen
  \bibfield  {author} {\bibinfo {author} {\bibfnamefont {P.~S.}\ \bibnamefont {Damle}}, \bibinfo {author} {\bibfnamefont {A.~W.}\ \bibnamefont {Ghosh}},\ and\ \bibinfo {author} {\bibfnamefont {S.}~\bibnamefont {Datta}},\ }\bibfield  {title} {\bibinfo {title} {Unified description of molecular conduction: From molecules to metallic wires},\ }\href {https://doi.org/10.1103/PhysRevB.64.201403} {\bibfield  {journal} {\bibinfo  {journal} {Phys. Rev. B}\ }\textbf {\bibinfo {volume} {64}},\ \bibinfo {pages} {201403} (\bibinfo {year} {2001})}\BibitemShut {NoStop}%
\bibitem [{\citenamefont {Brandbyge}\ \emph {et~al.}(2002)\citenamefont {Brandbyge}, \citenamefont {Mozos}, \citenamefont {Ordej\'on}, \citenamefont {Taylor},\ and\ \citenamefont {Stokbro}}]{Brandbyge2002}%
  \BibitemOpen
  \bibfield  {author} {\bibinfo {author} {\bibfnamefont {M.}~\bibnamefont {Brandbyge}}, \bibinfo {author} {\bibfnamefont {J.-L.}\ \bibnamefont {Mozos}}, \bibinfo {author} {\bibfnamefont {P.}~\bibnamefont {Ordej\'on}}, \bibinfo {author} {\bibfnamefont {J.}~\bibnamefont {Taylor}},\ and\ \bibinfo {author} {\bibfnamefont {K.}~\bibnamefont {Stokbro}},\ }\bibfield  {title} {\bibinfo {title} {Density-functional method for nonequilibrium electron transport},\ }\href {https://doi.org/10.1103/PhysRevB.65.165401} {\bibfield  {journal} {\bibinfo  {journal} {Phys. Rev. B}\ }\textbf {\bibinfo {volume} {65}},\ \bibinfo {pages} {165401} (\bibinfo {year} {2002})}\BibitemShut {NoStop}%
\bibitem [{\citenamefont {Xue}\ \emph {et~al.}(2002)\citenamefont {Xue}, \citenamefont {Datta},\ and\ \citenamefont {Ratner}}]{Xue2002}%
  \BibitemOpen
  \bibfield  {author} {\bibinfo {author} {\bibfnamefont {Y.}~\bibnamefont {Xue}}, \bibinfo {author} {\bibfnamefont {S.}~\bibnamefont {Datta}},\ and\ \bibinfo {author} {\bibfnamefont {M.~A.}\ \bibnamefont {Ratner}},\ }\bibfield  {title} {\bibinfo {title} {First-principles based matrix {G}reen's function approach to molecular electronic devices: general formalism},\ }\href {https://doi.org/https://doi.org/10.1016/S0301-0104(02)00446-9} {\bibfield  {journal} {\bibinfo  {journal} {Chemical Physics}\ }\textbf {\bibinfo {volume} {281}},\ \bibinfo {pages} {151} (\bibinfo {year} {2002})}\BibitemShut {NoStop}%
\bibitem [{\citenamefont {Camsari}\ \emph {et~al.}(2023)\citenamefont {Camsari}, \citenamefont {Chowdhury},\ and\ \citenamefont {Datta}}]{Camsari2023}%
  \BibitemOpen
  \bibfield  {author} {\bibinfo {author} {\bibfnamefont {K.}~\bibnamefont {Camsari}}, \bibinfo {author} {\bibfnamefont {S.}~\bibnamefont {Chowdhury}},\ and\ \bibinfo {author} {\bibfnamefont {S.}~\bibnamefont {Datta}},\ }\href@noop {} {\emph {\bibinfo {title} {Springer Handbook of Semiconductor Devices}}}\ (\bibinfo  {publisher} {Springer Cham},\ \bibinfo {year} {2023})\ p.\ \bibinfo {pages} {1583–1599}\BibitemShut {NoStop}%
\bibitem [{\citenamefont {Bertsch}\ and\ \citenamefont {Hagino}(2022)}]{Bertsch2022}%
  \BibitemOpen
  \bibfield  {author} {\bibinfo {author} {\bibfnamefont {G.~F.}\ \bibnamefont {Bertsch}}\ and\ \bibinfo {author} {\bibfnamefont {K.}~\bibnamefont {Hagino}},\ }\bibfield  {title} {\bibinfo {title} {Generator coordinate method for transition-state dynamics in nuclear fission},\ }\href {https://doi.org/10.1103/PhysRevC.105.034618} {\bibfield  {journal} {\bibinfo  {journal} {Phys. Rev. C}\ }\textbf {\bibinfo {volume} {105}},\ \bibinfo {pages} {034618} (\bibinfo {year} {2022})}\BibitemShut {NoStop}%
\bibitem [{\citenamefont {Weidenm\"uller}(2022)}]{Weidenmuller2022}%
  \BibitemOpen
  \bibfield  {author} {\bibinfo {author} {\bibfnamefont {H.~A.}\ \bibnamefont {Weidenm\"uller}},\ }\bibfield  {title} {\bibinfo {title} {Random-matrix approach to transition-state theory},\ }\href {https://doi.org/10.1103/PhysRevE.105.044143} {\bibfield  {journal} {\bibinfo  {journal} {Phys. Rev. E}\ }\textbf {\bibinfo {volume} {105}},\ \bibinfo {pages} {044143} (\bibinfo {year} {2022})}\BibitemShut {NoStop}%
\bibitem [{\citenamefont {Bertsch}\ and\ \citenamefont {Hagino}(2023)}]{Bertsch2023}%
  \BibitemOpen
  \bibfield  {author} {\bibinfo {author} {\bibfnamefont {G.~F.}\ \bibnamefont {Bertsch}}\ and\ \bibinfo {author} {\bibfnamefont {K.}~\bibnamefont {Hagino}},\ }\bibfield  {title} {\bibinfo {title} {Modeling fission dynamics at the barrier in a discrete-basis formalism},\ }\href {https://doi.org/10.1103/PhysRevC.107.044615} {\bibfield  {journal} {\bibinfo  {journal} {Phys. Rev. C}\ }\textbf {\bibinfo {volume} {107}},\ \bibinfo {pages} {044615} (\bibinfo {year} {2023})}\BibitemShut {NoStop}%
\bibitem [{\citenamefont {Uzawa}\ and\ \citenamefont {Hagino}(2023)}]{Uzawa2023}%
  \BibitemOpen
  \bibfield  {author} {\bibinfo {author} {\bibfnamefont {K.}~\bibnamefont {Uzawa}}\ and\ \bibinfo {author} {\bibfnamefont {K.}~\bibnamefont {Hagino}},\ }\bibfield  {title} {\bibinfo {title} {Schematic model for induced fission in a configuration-interaction approach},\ }\href {https://doi.org/10.1103/PhysRevC.108.024319} {\bibfield  {journal} {\bibinfo  {journal} {Phys. Rev. C}\ }\textbf {\bibinfo {volume} {108}},\ \bibinfo {pages} {024319} (\bibinfo {year} {2023})}\BibitemShut {NoStop}%
\bibitem [{\citenamefont {Uzawa}\ and\ \citenamefont {Hagino}(2024{\natexlab{a}})}]{Uzawa2024}%
  \BibitemOpen
  \bibfield  {author} {\bibinfo {author} {\bibfnamefont {K.}~\bibnamefont {Uzawa}}\ and\ \bibinfo {author} {\bibfnamefont {K.}~\bibnamefont {Hagino}},\ }\bibfield  {title} {\bibinfo {title} {Nonequilibrium {G}reen's function approach to low-energy fission dynamics: Fluctuations in fission reactions},\ }\href {https://doi.org/10.1103/PhysRevC.110.014321} {\bibfield  {journal} {\bibinfo  {journal} {Phys. Rev. C}\ }\textbf {\bibinfo {volume} {110}},\ \bibinfo {pages} {014321} (\bibinfo {year} {2024}{\natexlab{a}})}\BibitemShut {NoStop}%
\bibitem [{\citenamefont {Uzawa}\ and\ \citenamefont {Hagino}(2024{\natexlab{b}})}]{Uzawa2024-2}%
  \BibitemOpen
  \bibfield  {author} {\bibinfo {author} {\bibfnamefont {K.}~\bibnamefont {Uzawa}}\ and\ \bibinfo {author} {\bibfnamefont {K.}~\bibnamefont {Hagino}},\ }\bibfield  {title} {\bibinfo {title} {Application of the shift-invert {L}anczos algorithm to a nonequilibrium {G}reen's function for transport problems},\ }\href {https://doi.org/10.1103/PhysRevE.110.055302} {\bibfield  {journal} {\bibinfo  {journal} {Phys. Rev. E}\ }\textbf {\bibinfo {volume} {110}},\ \bibinfo {pages} {055302} (\bibinfo {year} {2024}{\natexlab{b}})}\BibitemShut {NoStop}%
\bibitem [{\citenamefont {Moore}\ \emph {et~al.}(1984)\citenamefont {Moore}, \citenamefont {Calabretta}, \citenamefont {Corvi},\ and\ \citenamefont {Weigmann}}]{Moore1984}%
  \BibitemOpen
  \bibfield  {author} {\bibinfo {author} {\bibfnamefont {M.~S.}\ \bibnamefont {Moore}}, \bibinfo {author} {\bibfnamefont {L.}~\bibnamefont {Calabretta}}, \bibinfo {author} {\bibfnamefont {F.}~\bibnamefont {Corvi}},\ and\ \bibinfo {author} {\bibfnamefont {H.}~\bibnamefont {Weigmann}},\ }\bibfield  {title} {\bibinfo {title} {Analysis of intermediate structure in the fission and capture cross sections of ($^{235}\mathrm{U}$+n)},\ }\href {https://doi.org/10.1103/PhysRevC.30.214} {\bibfield  {journal} {\bibinfo  {journal} {Phys. Rev. C}\ }\textbf {\bibinfo {volume} {30}},\ \bibinfo {pages} {214} (\bibinfo {year} {1984})}\BibitemShut {NoStop}%
\bibitem [{\citenamefont {Bertsch}\ and\ \citenamefont {Kawano}(2017)}]{Bertsch2017}%
  \BibitemOpen
  \bibfield  {author} {\bibinfo {author} {\bibfnamefont {G.~F.}\ \bibnamefont {Bertsch}}\ and\ \bibinfo {author} {\bibfnamefont {T.}~\bibnamefont {Kawano}},\ }\bibfield  {title} {\bibinfo {title} {Exit-channel suppression in statistical reaction theory},\ }\href {https://doi.org/10.1103/PhysRevLett.119.222504} {\bibfield  {journal} {\bibinfo  {journal} {Phys. Rev. Lett.}\ }\textbf {\bibinfo {volume} {119}},\ \bibinfo {pages} {222504} (\bibinfo {year} {2017})}\BibitemShut {NoStop}%
\bibitem [{\citenamefont {Kortelainen}\ \emph {et~al.}(2012)\citenamefont {Kortelainen}, \citenamefont {McDonnell}, \citenamefont {Nazarewicz}, \citenamefont {Reinhard}, \citenamefont {Sarich}, \citenamefont {Schunck}, \citenamefont {Stoitsov},\ and\ \citenamefont {Wild}}]{Kortelainen2012}%
  \BibitemOpen
  \bibfield  {author} {\bibinfo {author} {\bibfnamefont {M.}~\bibnamefont {Kortelainen}}, \bibinfo {author} {\bibfnamefont {J.}~\bibnamefont {McDonnell}}, \bibinfo {author} {\bibfnamefont {W.}~\bibnamefont {Nazarewicz}}, \bibinfo {author} {\bibfnamefont {P.-G.}\ \bibnamefont {Reinhard}}, \bibinfo {author} {\bibfnamefont {J.}~\bibnamefont {Sarich}}, \bibinfo {author} {\bibfnamefont {N.}~\bibnamefont {Schunck}}, \bibinfo {author} {\bibfnamefont {M.~V.}\ \bibnamefont {Stoitsov}},\ and\ \bibinfo {author} {\bibfnamefont {S.~M.}\ \bibnamefont {Wild}},\ }\bibfield  {title} {\bibinfo {title} {Nuclear energy density optimization: Large deformations},\ }\href {https://doi.org/10.1103/PhysRevC.85.024304} {\bibfield  {journal} {\bibinfo  {journal} {Phys. Rev. C}\ }\textbf {\bibinfo {volume} {85}},\ \bibinfo {pages} {024304} (\bibinfo {year} {2012})}\BibitemShut {NoStop}%
\bibitem [{\citenamefont {Ring}\ and\ \citenamefont {Schuck}(2000)}]{ring}%
  \BibitemOpen
  \bibfield  {author} {\bibinfo {author} {\bibfnamefont {P.}~\bibnamefont {Ring}}\ and\ \bibinfo {author} {\bibfnamefont {P.}~\bibnamefont {Schuck}},\ }\href@noop {} {\emph {\bibinfo {title} {The Nuclear Many-Body Problem}}}\ (\bibinfo  {publisher} {Springer-Verlag, Berlin},\ \bibinfo {year} {2000})\BibitemShut {NoStop}%
\bibitem [{\citenamefont {Reinhard}\ \emph {et~al.}(2021)\citenamefont {Reinhard}, \citenamefont {Schuetrumpf},\ and\ \citenamefont {Maruhn}}]{Reinhard2021}%
  \BibitemOpen
  \bibfield  {author} {\bibinfo {author} {\bibfnamefont {P.-G.}\ \bibnamefont {Reinhard}}, \bibinfo {author} {\bibfnamefont {B.}~\bibnamefont {Schuetrumpf}},\ and\ \bibinfo {author} {\bibfnamefont {J.}~\bibnamefont {Maruhn}},\ }\bibfield  {title} {\bibinfo {title} {The {A}xial {H}artree–{F}ock + {BCS} {C}ode {S}ky{A}x},\ }\href {https://doi.org/https://doi.org/10.1016/j.cpc.2020.107603} {\bibfield  {journal} {\bibinfo  {journal} {Comput. Phys. Commun.}\ }\textbf {\bibinfo {volume} {258}},\ \bibinfo {pages} {107603} (\bibinfo {year} {2021})}\BibitemShut {NoStop}%
\bibitem [{\citenamefont {Leal}\ \emph {et~al.}(1999)\citenamefont {Leal}, \citenamefont {Derrien}, \citenamefont {Larson},\ and\ \citenamefont {Wright}}]{Leal1999}%
  \BibitemOpen
  \bibfield  {author} {\bibinfo {author} {\bibfnamefont {L.~C.}\ \bibnamefont {Leal}}, \bibinfo {author} {\bibfnamefont {H.}~\bibnamefont {Derrien}}, \bibinfo {author} {\bibfnamefont {N.~M.}\ \bibnamefont {Larson}},\ and\ \bibinfo {author} {\bibfnamefont {R.~Q.}\ \bibnamefont {Wright}},\ }\bibfield  {title} {\bibinfo {title} {R-{M}atrix {A}nalysis of $^{235}${U} {N}eutron {T}ransmission and {C}ross-{S}ection {M}easurements in the 0- to 2.25-ke{V} {E}nergy {R}ange},\ }\href {https://doi.org/10.13182/NSE99-A2031} {\bibfield  {journal} {\bibinfo  {journal} {Nucl. Sci. Eng.}\ }\textbf {\bibinfo {volume} {131}},\ \bibinfo {pages} {230} (\bibinfo {year} {1999})}\BibitemShut {NoStop}%
\bibitem [{\citenamefont {Hagino}\ and\ \citenamefont {Bertsch}(2022)}]{Hagino2022}%
  \BibitemOpen
  \bibfield  {author} {\bibinfo {author} {\bibfnamefont {K.}~\bibnamefont {Hagino}}\ and\ \bibinfo {author} {\bibfnamefont {G.~F.}\ \bibnamefont {Bertsch}},\ }\bibfield  {title} {\bibinfo {title} {Diabatic hamiltonian matrix elements made simple},\ }\href {https://doi.org/10.1103/PhysRevC.105.034323} {\bibfield  {journal} {\bibinfo  {journal} {Phys. Rev. C}\ }\textbf {\bibinfo {volume} {105}},\ \bibinfo {pages} {034323} (\bibinfo {year} {2022})}\BibitemShut {NoStop}%
\bibitem [{\citenamefont {Bush}\ \emph {et~al.}(1992)\citenamefont {Bush}, \citenamefont {Bertsch},\ and\ \citenamefont {Brown}}]{Bush1992}%
  \BibitemOpen
  \bibfield  {author} {\bibinfo {author} {\bibfnamefont {B.~W.}\ \bibnamefont {Bush}}, \bibinfo {author} {\bibfnamefont {G.~F.}\ \bibnamefont {Bertsch}},\ and\ \bibinfo {author} {\bibfnamefont {B.~A.}\ \bibnamefont {Brown}},\ }\bibfield  {title} {\bibinfo {title} {Shape diffusion in the shell model},\ }\href {https://doi.org/10.1103/PhysRevC.45.1709} {\bibfield  {journal} {\bibinfo  {journal} {Phys. Rev. C}\ }\textbf {\bibinfo {volume} {45}},\ \bibinfo {pages} {1709} (\bibinfo {year} {1992})}\BibitemShut {NoStop}%
\bibitem [{\citenamefont {Capote}\ \emph {et~al.}(2009)\citenamefont {Capote}, \citenamefont {Herman}, \citenamefont {Obložinský}, \citenamefont {Young}, \citenamefont {Goriely}, \citenamefont {Belgya}, \citenamefont {Ignatyuk}, \citenamefont {Koning}, \citenamefont {Hilaire}, \citenamefont {Plujko}, \citenamefont {Avrigeanu}, \citenamefont {Bersillon}, \citenamefont {Chadwick}, \citenamefont {Fukahori}, \citenamefont {Ge}, \citenamefont {Han}, \citenamefont {Kailas}, \citenamefont {Kopecky}, \citenamefont {Maslov}, \citenamefont {Reffo}, \citenamefont {Sin}, \citenamefont {Soukhovitskii},\ and\ \citenamefont {Talou}}]{ripl}%
  \BibitemOpen
  \bibfield  {author} {\bibinfo {author} {\bibfnamefont {R.}~\bibnamefont {Capote}}, \bibinfo {author} {\bibfnamefont {M.}~\bibnamefont {Herman}}, \bibinfo {author} {\bibfnamefont {P.}~\bibnamefont {Obložinský}}, \bibinfo {author} {\bibfnamefont {P.}~\bibnamefont {Young}}, \bibinfo {author} {\bibfnamefont {S.}~\bibnamefont {Goriely}}, \bibinfo {author} {\bibfnamefont {T.}~\bibnamefont {Belgya}}, \bibinfo {author} {\bibfnamefont {A.}~\bibnamefont {Ignatyuk}}, \bibinfo {author} {\bibfnamefont {A.}~\bibnamefont {Koning}}, \bibinfo {author} {\bibfnamefont {S.}~\bibnamefont {Hilaire}}, \bibinfo {author} {\bibfnamefont {V.}~\bibnamefont {Plujko}}, \bibinfo {author} {\bibfnamefont {M.}~\bibnamefont {Avrigeanu}}, \bibinfo {author} {\bibfnamefont {O.}~\bibnamefont {Bersillon}}, \bibinfo {author} {\bibfnamefont {M.}~\bibnamefont {Chadwick}}, \bibinfo {author} {\bibfnamefont {T.}~\bibnamefont {Fukahori}}, \bibinfo {author} {\bibfnamefont {Z.}~\bibnamefont {Ge}}, \bibinfo {author} {\bibfnamefont {Y.}~\bibnamefont
  {Han}}, \bibinfo {author} {\bibfnamefont {S.}~\bibnamefont {Kailas}}, \bibinfo {author} {\bibfnamefont {J.}~\bibnamefont {Kopecky}}, \bibinfo {author} {\bibfnamefont {V.}~\bibnamefont {Maslov}}, \bibinfo {author} {\bibfnamefont {G.}~\bibnamefont {Reffo}}, \bibinfo {author} {\bibfnamefont {M.}~\bibnamefont {Sin}}, \bibinfo {author} {\bibfnamefont {E.}~\bibnamefont {Soukhovitskii}},\ and\ \bibinfo {author} {\bibfnamefont {P.}~\bibnamefont {Talou}},\ }\bibfield  {title} {\bibinfo {title} {{RIPL} – reference input parameter library for calculation of nuclear reactions and nuclear data evaluations},\ }\href {https://doi.org/https://doi.org/10.1016/j.nds.2009.10.004} {\bibfield  {journal} {\bibinfo  {journal} {Nuclear Data Sheets}\ }\textbf {\bibinfo {volume} {110}},\ \bibinfo {pages} {3107} (\bibinfo {year} {2009})},\ \bibinfo {note} {special Issue on Nuclear Reaction Data}\BibitemShut {NoStop}%
\bibitem [{\citenamefont {Brody}\ \emph {et~al.}(1981)\citenamefont {Brody}, \citenamefont {Flores}, \citenamefont {French}, \citenamefont {Mello}, \citenamefont {Pandey},\ and\ \citenamefont {Wong}}]{Brody1981}%
  \BibitemOpen
  \bibfield  {author} {\bibinfo {author} {\bibfnamefont {T.~A.}\ \bibnamefont {Brody}}, \bibinfo {author} {\bibfnamefont {J.}~\bibnamefont {Flores}}, \bibinfo {author} {\bibfnamefont {J.~B.}\ \bibnamefont {French}}, \bibinfo {author} {\bibfnamefont {P.~A.}\ \bibnamefont {Mello}}, \bibinfo {author} {\bibfnamefont {A.}~\bibnamefont {Pandey}},\ and\ \bibinfo {author} {\bibfnamefont {S.~S.~M.}\ \bibnamefont {Wong}},\ }\bibfield  {title} {\bibinfo {title} {Random-matrix physics: spectrum and strength fluctuations},\ }\href {https://doi.org/10.1103/RevModPhys.53.385} {\bibfield  {journal} {\bibinfo  {journal} {Rev. Mod. Phys.}\ }\textbf {\bibinfo {volume} {53}},\ \bibinfo {pages} {385} (\bibinfo {year} {1981})}\BibitemShut {NoStop}%
\bibitem [{\citenamefont {Bertsch}\ and\ \citenamefont {Robledo}(2019)}]{Bertsch2019}%
  \BibitemOpen
  \bibfield  {author} {\bibinfo {author} {\bibfnamefont {G.~F.}\ \bibnamefont {Bertsch}}\ and\ \bibinfo {author} {\bibfnamefont {L.~M.}\ \bibnamefont {Robledo}},\ }\bibfield  {title} {\bibinfo {title} {Decay widths at the scission point in nuclear fission},\ }\href {https://doi.org/10.1103/PhysRevC.100.044606} {\bibfield  {journal} {\bibinfo  {journal} {Phys. Rev. C}\ }\textbf {\bibinfo {volume} {100}},\ \bibinfo {pages} {044606} (\bibinfo {year} {2019})}\BibitemShut {NoStop}%
\bibitem [{\citenamefont {Caroli}\ \emph {et~al.}(1971)\citenamefont {Caroli}, \citenamefont {Combescot}, \citenamefont {Nozieres},\ and\ \citenamefont {Saint-James}}]{Caroli1971}%
  \BibitemOpen
  \bibfield  {author} {\bibinfo {author} {\bibfnamefont {C.}~\bibnamefont {Caroli}}, \bibinfo {author} {\bibfnamefont {R.}~\bibnamefont {Combescot}}, \bibinfo {author} {\bibfnamefont {P.}~\bibnamefont {Nozieres}},\ and\ \bibinfo {author} {\bibfnamefont {D.}~\bibnamefont {Saint-James}},\ }\bibfield  {title} {\bibinfo {title} {Direct calculation of the tunneling current},\ }\href {https://doi.org/10.1088/0022-3719/4/8/018} {\bibfield  {journal} {\bibinfo  {journal} {Journal of Physics C: Solid State Physics}\ }\textbf {\bibinfo {volume} {4}},\ \bibinfo {pages} {916} (\bibinfo {year} {1971})}\BibitemShut {NoStop}%
\bibitem [{\citenamefont {Meir}\ and\ \citenamefont {Wingreen}(1992)}]{Meir1992}%
  \BibitemOpen
  \bibfield  {author} {\bibinfo {author} {\bibfnamefont {Y.}~\bibnamefont {Meir}}\ and\ \bibinfo {author} {\bibfnamefont {N.~S.}\ \bibnamefont {Wingreen}},\ }\bibfield  {title} {\bibinfo {title} {{L}andauer formula for the current through an interacting electron region},\ }\href {https://doi.org/10.1103/PhysRevLett.68.2512} {\bibfield  {journal} {\bibinfo  {journal} {Phys. Rev. Lett.}\ }\textbf {\bibinfo {volume} {68}},\ \bibinfo {pages} {2512} (\bibinfo {year} {1992})}\BibitemShut {NoStop}%
\bibitem [{\citenamefont {Alhassid}\ \emph {et~al.}(2021)\citenamefont {Alhassid}, \citenamefont {Bertsch},\ and\ \citenamefont {Fanto}}]{Alhassid2021}%
  \BibitemOpen
  \bibfield  {author} {\bibinfo {author} {\bibfnamefont {Y.}~\bibnamefont {Alhassid}}, \bibinfo {author} {\bibfnamefont {G.~F.}\ \bibnamefont {Bertsch}},\ and\ \bibinfo {author} {\bibfnamefont {P.}~\bibnamefont {Fanto}},\ }\bibfield  {title} {\bibinfo {title} {Addendum to “derivation of {K}-matrix reaction theory in a discrete basis formalism” [ann. phys. 419 (2020) 168233]},\ }\href {https://doi.org/https://doi.org/10.1016/j.aop.2020.168381} {\bibfield  {journal} {\bibinfo  {journal} {Annals of Physics}\ }\textbf {\bibinfo {volume} {424}},\ \bibinfo {pages} {168381} (\bibinfo {year} {2021})}\BibitemShut {NoStop}%
\bibitem [{\citenamefont {Porter}\ and\ \citenamefont {Thomas}(1956)}]{Porter1956}%
  \BibitemOpen
  \bibfield  {author} {\bibinfo {author} {\bibfnamefont {C.~E.}\ \bibnamefont {Porter}}\ and\ \bibinfo {author} {\bibfnamefont {R.~G.}\ \bibnamefont {Thomas}},\ }\bibfield  {title} {\bibinfo {title} {Fluctuations of nuclear reaction widths},\ }\href {https://doi.org/10.1103/PhysRev.104.483} {\bibfield  {journal} {\bibinfo  {journal} {Phys. Rev.}\ }\textbf {\bibinfo {volume} {104}},\ \bibinfo {pages} {483} (\bibinfo {year} {1956})}\BibitemShut {NoStop}%
\bibitem [{\citenamefont {Huang}\ \emph {et~al.}(2012)\citenamefont {Huang}, \citenamefont {Wang}, \citenamefont {Yan},\ and\ \citenamefont {Chi}}]{LSMR}%
  \BibitemOpen
  \bibfield  {author} {\bibinfo {author} {\bibfnamefont {Q.-X.}\ \bibnamefont {Huang}}, \bibinfo {author} {\bibfnamefont {F.}~\bibnamefont {Wang}}, \bibinfo {author} {\bibfnamefont {J.-H.}\ \bibnamefont {Yan}, \bibfnamefont {Fei}},\ and\ \bibinfo {author} {\bibfnamefont {Y.}~\bibnamefont {Chi}},\ }\bibfield  {title} {\bibinfo {title} {A two-step discrete method for reconstruction of temperature distribution in a three-dimensional participating medium},\ }\href {https://doi.org/https://doi.org/10.1016/j.ijheatmasstransfer.2011.12.029} {\bibfield  {journal} {\bibinfo  {journal} {International Journal of Heat and Mass Transfer}\ }\textbf {\bibinfo {volume} {55}},\ \bibinfo {pages} {2636} (\bibinfo {year} {2012})}\BibitemShut {NoStop}%
\bibitem [{\citenamefont {Iwamoto}\ \emph {et~al.}(2023)\citenamefont {Iwamoto}, \citenamefont {Iwamoto}, \citenamefont {Kunieda}, \citenamefont {Minato}, \citenamefont {Nakayama}, \citenamefont {Abe}, \citenamefont {Tsubakihara}, \citenamefont {Okumura}, \citenamefont {Ishizuka}, \citenamefont {Yoshida} \emph {et~al.}}]{Iwamoto2023}%
  \BibitemOpen
  \bibfield  {author} {\bibinfo {author} {\bibfnamefont {O.}~\bibnamefont {Iwamoto}}, \bibinfo {author} {\bibfnamefont {N.}~\bibnamefont {Iwamoto}}, \bibinfo {author} {\bibfnamefont {S.}~\bibnamefont {Kunieda}}, \bibinfo {author} {\bibfnamefont {F.}~\bibnamefont {Minato}}, \bibinfo {author} {\bibfnamefont {S.}~\bibnamefont {Nakayama}}, \bibinfo {author} {\bibfnamefont {Y.}~\bibnamefont {Abe}}, \bibinfo {author} {\bibfnamefont {K.}~\bibnamefont {Tsubakihara}}, \bibinfo {author} {\bibfnamefont {S.}~\bibnamefont {Okumura}}, \bibinfo {author} {\bibfnamefont {C.}~\bibnamefont {Ishizuka}}, \bibinfo {author} {\bibfnamefont {T.}~\bibnamefont {Yoshida}}, \emph {et~al.},\ }\bibfield  {title} {\bibinfo {title} {Japanese evaluated nuclear data library version 5: {JENDL}-5},\ }\href@noop {} {\bibfield  {journal} {\bibinfo  {journal} {Journal of Nuclear Science and Technology}\ }\textbf {\bibinfo {volume} {60}},\ \bibinfo {pages} {1} (\bibinfo {year} {2023})}\BibitemShut {NoStop}%
\bibitem [{\citenamefont {Goeke}\ and\ \citenamefont {Reinhard}(1980)}]{Goeke1980}%
  \BibitemOpen
  \bibfield  {author} {\bibinfo {author} {\bibfnamefont {K.}~\bibnamefont {Goeke}}\ and\ \bibinfo {author} {\bibfnamefont {P.-G.}\ \bibnamefont {Reinhard}},\ }\bibfield  {title} {\bibinfo {title} {The generator-coordinate-method with conjugate parameters and the unification of microscopic theories for large amplitude collective motion},\ }\href {https://doi.org/https://doi.org/10.1016/0003-4916(80)90210-9} {\bibfield  {journal} {\bibinfo  {journal} {Annals of Physics}\ }\textbf {\bibinfo {volume} {124}},\ \bibinfo {pages} {249} (\bibinfo {year} {1980})}\BibitemShut {NoStop}%
\bibitem [{\citenamefont {Hizawa}\ \emph {et~al.}(2022)\citenamefont {Hizawa}, \citenamefont {Hagino},\ and\ \citenamefont {Yoshida}}]{Hizawa2022}%
  \BibitemOpen
  \bibfield  {author} {\bibinfo {author} {\bibfnamefont {N.}~\bibnamefont {Hizawa}}, \bibinfo {author} {\bibfnamefont {K.}~\bibnamefont {Hagino}},\ and\ \bibinfo {author} {\bibfnamefont {K.}~\bibnamefont {Yoshida}},\ }\bibfield  {title} {\bibinfo {title} {Applications of the dynamical generator coordinate method to quadrupole excitations},\ }\href {https://doi.org/10.1103/PhysRevC.105.064302} {\bibfield  {journal} {\bibinfo  {journal} {Phys. Rev. C}\ }\textbf {\bibinfo {volume} {105}},\ \bibinfo {pages} {064302} (\bibinfo {year} {2022})}\BibitemShut {NoStop}%
\bibitem [{\citenamefont {Hagino}\ and\ \citenamefont {Bertsch}(2024)}]{Hagino2024}%
  \BibitemOpen
  \bibfield  {author} {\bibinfo {author} {\bibfnamefont {K.}~\bibnamefont {Hagino}}\ and\ \bibinfo {author} {\bibfnamefont {G.~F.}\ \bibnamefont {Bertsch}},\ }\bibfield  {title} {\bibinfo {title} {Role of momentum in the generator-coordinate method applied to barrier penetration},\ }\href {https://doi.org/10.1103/PhysRevC.110.054610} {\bibfield  {journal} {\bibinfo  {journal} {Phys. Rev. C}\ }\textbf {\bibinfo {volume} {110}},\ \bibinfo {pages} {054610} (\bibinfo {year} {2024})}\BibitemShut {NoStop}%
\bibitem [{\citenamefont {Ericsson}\ and\ \citenamefont {Ruhe}(1980)}]{Shift-Invert}%
  \BibitemOpen
  \bibfield  {author} {\bibinfo {author} {\bibfnamefont {T.}~\bibnamefont {Ericsson}}\ and\ \bibinfo {author} {\bibfnamefont {A.}~\bibnamefont {Ruhe}},\ }\bibfield  {title} {\bibinfo {title} {The spectral transformation lanczos method for the numerical solution of large sparse generalized symmetric eigenvalue problems},\ }\href {http://www.jstor.org/stable/2006390} {\bibfield  {journal} {\bibinfo  {journal} {Mathematics of Computation}\ }\textbf {\bibinfo {volume} {35}},\ \bibinfo {pages} {1251} (\bibinfo {year} {1980})}\BibitemShut {NoStop}%
\bibitem [{\citenamefont {Chabanat}\ \emph {et~al.}(1997)\citenamefont {Chabanat}, \citenamefont {Bonche}, \citenamefont {Haensel}, \citenamefont {Meyer},\ and\ \citenamefont {Schaeffer}}]{CHABANAT1997710}%
  \BibitemOpen
  \bibfield  {author} {\bibinfo {author} {\bibfnamefont {E.}~\bibnamefont {Chabanat}}, \bibinfo {author} {\bibfnamefont {P.}~\bibnamefont {Bonche}}, \bibinfo {author} {\bibfnamefont {P.}~\bibnamefont {Haensel}}, \bibinfo {author} {\bibfnamefont {J.}~\bibnamefont {Meyer}},\ and\ \bibinfo {author} {\bibfnamefont {R.}~\bibnamefont {Schaeffer}},\ }\bibfield  {title} {\bibinfo {title} {A {S}kyrme parametrization from subnuclear to neutron star densities},\ }\href {https://doi.org/https://doi.org/10.1016/S0375-9474(97)00596-4} {\bibfield  {journal} {\bibinfo  {journal} {Nuclear Physics A}\ }\textbf {\bibinfo {volume} {627}},\ \bibinfo {pages} {710} (\bibinfo {year} {1997})}\BibitemShut {NoStop}%
\bibitem [{\citenamefont {Bartel}\ \emph {et~al.}(1982)\citenamefont {Bartel}, \citenamefont {Quentin}, \citenamefont {Brack}, \citenamefont {Guet},\ and\ \citenamefont {Håkansson}}]{BARTEL198279}%
  \BibitemOpen
  \bibfield  {author} {\bibinfo {author} {\bibfnamefont {J.}~\bibnamefont {Bartel}}, \bibinfo {author} {\bibfnamefont {P.}~\bibnamefont {Quentin}}, \bibinfo {author} {\bibfnamefont {M.}~\bibnamefont {Brack}}, \bibinfo {author} {\bibfnamefont {C.}~\bibnamefont {Guet}},\ and\ \bibinfo {author} {\bibfnamefont {H.-B.}\ \bibnamefont {Håkansson}},\ }\bibfield  {title} {\bibinfo {title} {Towards a better parametrisation of {S}kyrme-like effective forces: A critical study of the {S}k{M} force},\ }\href {https://doi.org/https://doi.org/10.1016/0375-9474(82)90403-1} {\bibfield  {journal} {\bibinfo  {journal} {Nuclear Physics A}\ }\textbf {\bibinfo {volume} {386}},\ \bibinfo {pages} {79} (\bibinfo {year} {1982})}\BibitemShut {NoStop}%
\end{thebibliography}

\end{document}